\documentclass[journal,twoside,10pt]{IEEEtran}
\usepackage{etoolbox}
\usepackage{soul, xcolor, comment}  %
\usepackage{amsmath,amsfonts}
\usepackage{algorithmic,algorithm}
\usepackage{float}
\usepackage{graphicx} %
\usepackage[caption=false,font=normalsize,labelfont=sf,textfont=sf]{subfig}
\usepackage[absolute,overlay]{textpos} %
\usepackage{textcomp}
\usepackage{stfloats}
\usepackage{url}
\usepackage{verbatim}
\usepackage{enumitem} %
\usepackage{array,booktabs,multirow,tabularx,makecell,arydshln} %

\usepackage[numbers,sort&compress]{natbib}

\usepackage[breaklinks,colorlinks]{hyperref}
\usepackage[capitalize]{cleveref}
\usepackage{needspace} %

\newcommand{\customsection}[1]{\needspace{3\baselineskip}\vspace{1em}\noindent \textbf{\textit{#1}}\vspace{0.25em}}
\Crefname{section}{Section}{Sections}
\crefname{section}{Sec.}{Secs.}
\Crefname{table}{Table}{Tables}
\crefname{table}{Tab.}{Tabs.}
\Crefname{figure}{Figure}{Figures}
\crefname{figure}{Fig.}{Figs.}

\graphicspath{{./figures/}{./portraits}}

\let\hlold\hl
\renewcommand{\hl}[1]{\textcolor{red}{\hlold{#1}}}

\begin{document}

\author{
Jinchao Li$^{*}$,
Yuejiao Wang$^{*}$,
Junan Li$^{*}$,
Jiawen Kang$^{*}$,
Bo Zheng,
Ka Ho Wong,\\
Brian Mak,
Helene H. Fung,
Jean Woo,
Man-Wai Mak,
Timothy Kwok,
Vincent Mok,\\
Xianmin Gong,
Xixin Wu,
Xunying Liu,
Patrick C. M. Wong,
Helen Meng,~\IEEEmembership{Fellow,~IEEE}

\thanks{
$^{*}$These authors contributed equally to this work (e-mails: \{{\tt jcli}, {\tt wangy}, {\tt jli}, {\tt jwkang}\}{\tt @se.cuhk.edu.hk}).
Jinchao Li, Yuejiao Wang, Junan Li,
Jiawen Kang, Bo Zheng, Ka Ho Wong, Helene H. Fung, Jean Woo, Timothy Kwok, Vincent Mok, Xianmin Gong, Xixin Wu, Xunying Liu, Patrick C. M. Wong, and Helen Meng (e-mail: {\tt hmmeng@se.cuhk.edu.hk}) are with The Chinese University of Hong Kong, Hong Kong, China.
Man-Wai Mak is with The Hong Kong Polytechnic University, Hong Kong, China.
Brian Mak is with The Hong Kong University of Science and Technology, Hong Kong, China.
This project is partially supported by the HKSAR Research Grants Council (Project No. T45-407/19N).
% 
% Digital Object Identifier: 10.1109/JSTSP.2025.3622049
}

}

\title{%
Detecting Neurocognitive Disorders through Analyses of Topic Evolution and Cross-Modal Consistency in Visual-Stimulated Narratives
}
\maketitle

\markboth{IEEE Journal of Selected Topics in Signal Processing,~Vol.~X, No.~Y,~Month~Year}%
{Li \MakeLowercase{\textit{et al.}}:
Detecting NCDs through Analyses of Topic Evolution and Cross-modal Consistency in Visual-Stimulated Narratives
}

\IEEEpubid{0000--0000/00\$00.00~\copyright~YEAR IEEE}
\IEEEpubidadjcol

\begin{abstract}

Early detection of neurocognitive disorders (NCDs) is crucial for timely intervention and disease management. Given that language impairments manifest early in NCD progression, visual-stimulated narrative (VSN)-based analysis offers a promising avenue for NCD detection. Current VSN-based NCD detection methods primarily focus on linguistic microstructures (e.g., lexical diversity) that are closely tied to bottom-up, stimulus-driven cognitive processes. While these features illuminate basic language abilities, the higher-order linguistic macrostructures (e.g., topic development) that may reflect top-down, concept-driven cognitive abilities remain underexplored. These macrostructural patterns are crucial for NCD detection, yet challenging to quantify due to their abstract and complex nature. To bridge this gap, we propose two novel macrostructural approaches: (1) a Dynamic Topic Model (DTM) to track topic evolution over time, and (2) a Text-Image Temporal Alignment Network (TITAN) to measure cross-modal consistency between narrative and visual stimuli. Experimental results show the effectiveness of the proposed approaches in NCD detection, with TITAN achieving superior performance across three corpora: ADReSS (F1=0.8889), ADReSSo (F1=0.8504), and CU-MARVEL-RABBIT (F1=0.7238). Feature contribution analysis reveals that macrostructural features (e.g., topic variability, topic change rate, and topic consistency) constitute the most significant contributors to the model's decision pathways, outperforming the investigated microstructural features. These findings underscore the value of macrostructural analysis for understanding linguistic-cognitive interactions associated with NCDs.

\end{abstract}


\section{Introduction}
\label{sec:intro}

Neurocognitive disorders (NCDs)\cite{american2013diagnostic,sachdev2014classifying}, such as Alzheimer's Disease, are highly prevalent and pose significant challenges to individuals and healthcare systems in our aging society.
Early detection of these conditions is critical, as it facilitates timely intervention and disease management~\cite{adi2024world}.
Traditional clinical approaches such as neuroimaging~\cite{wang2024naturalistic,wang2024large}, retinal photographs~\cite{ashayeri2024retinal}, electroencephalography (EEG)~\cite{siuly2024exploring}, cerebrospinal fluid (CSF) analysis~\cite{guo2024multiplex}, and in-person neuropsychological assessments~\cite{veneziani2024applications}, have been widely investigated.
However, these approaches are often resource-intensive, invasive, and costly~\cite{sharma2023deep,lim2023quantification}.
Consequently, there is increasing interest in developing non-invasive, cost-effective, and scalable methods for early NCD detection.

\IEEEpubidadjcol %

One promising approach is the automatic analysis of speech and language, as communicative impairments are often among the earliest symptoms of cognitive decline~\cite{wolf2024early}.
These impairments may include temporal changes in spontaneous speech~\cite{melistas23_interspeech}, word finding and retrieval difficulties~\cite{martinez2024naming}, verbal dysfluencies~\cite{wright2023current}, and structural and thematic organization difficulties~\cite{murray1999discourse,mueller2024using}.
Such changes in speech patterns and language use provide valuable insights into underlying cognitive deficits and serve as potential digital biomarkers for early NCD detection.
In light of this, various spoken cognitive tests, such as counting~\cite{logie1987cognitive} and naming~\cite{kaplan2001boston}, have been designed to identify spoken language correlates for NCD detection.
Among these, the visual-stimulated narrative (VSN) task~\cite{chenery1994production,croisile1996comparative}, such as describing a single image or telling a story based on image sequences, stands out due to its ability to elicit rich and spontaneous speech through visual stimuli within a naturalistic and holistic framework.
Specifically, the VSN task involves cognitive-linguistic processes such as visual perception, verbal description, event sequencing, and narrative integration, thereby engaging multiple cognitive domains, including attention, visual comprehension, language fluency, executive planning, memory, and social cognition~\cite{kintsch1978cognitive,marini2005age,Helimann2009NSS,labov2010oral,de2011microlinguistic,toledo2018analysis,yang2024makes}.
Therefore, it holds strong promise for comprehensively assessment of cognitive abilities for NCD detection~\cite{possin2010visual,wolf2024early}.

Approaches to the analysis of VSNs can be categorized into two distinct yet complementary types~\cite{kong2023spoken}.
First, the mainstream approach is \textbf{microstructural} analysis, which examines detailed, localized narrative elements, such as linguistic (e.g., lexicon, semantics, syntax)~\cite{li2021comparative,pl2024cognitive} and acoustic (e.g., prosody, phonology)~\cite{kurdi2024dementia,saeedi2024acoustic} units, primarily reflecting bottom-up (exogenous, stimulus-driven) cognitive processing.
In contrast, \textbf{macrostructural} analysis evaluates the overarching organization of the narrative units, such as informativeness (overall information accuracy and completeness), coherence (logical flow of events or concepts) and text-visual consistency (alignment with visual stimuli\footnote{This includes both depiction per image and connection across images, bridging informativeness and coherence.})~\cite{geelhand2020narrative}, generally representing top-down (endogenous, concept-driven) cognitive integration.

Clinical research indicates that individuals with NCDs often exhibit greater impairment in macrostructuring than microstructuring.
While microstructural deficits are associated with basic language abilities (perception and production), macrostructural deficits more directly reflect impairments in higher-order cognitive abilities (attention, memory, organization, executive function), manifesting as disrupted topic maintenance, incoherent sequencing, and inconsistency with visual stimuli~\cite{kong2023spoken}.
Despite this, macrostructural analysis remains underexplored in NCD narratives.

This research gap stems largely from the abstract and conceptual nature of macrostructural features, which makes them inherently difficult to quantify.
Traditional approaches to quantify coherence rely on manual assessments, which are subjective and labor-intensive~\cite{kokje2023macro}.
Recent advances in natural language processing (NLP) have introduced automated methods, such as computing cosine similarities between consecutive textual embeddings or employing large language models (LLMs) as evaluators~\cite{burke2023comparing,botelho2024macro}.
However, these methods often fall short for NCD narratives.
Embedding-based similarities may erroneously interpret frequent repetitions or fillers as signs of high cohesion, while failing to capture the dynamic evolution of narrative topics over time.
LLM-generated scores lack interpretability due to their untraceable decision pathways, undermining their reliability in clinical applications.

To quantify the consistency between narratives and visual stimuli, recent studies employed text-visual pretrained models to measure the highest relevance scores between image regions and sentences~\cite{zhu2023evaluating} or used pattern-matching methods to compute narrative similarity based on fixed references~\cite{li2023leveraging,li2024devising}.
However, these methods have limitations: local text-image relevance focuses on region-level consistency and fails to capture global narrative coherence or its temporal evolution; meanwhile, pattern-matching approaches depend on predefined references, rendering them inadequate for open-ended, creative tasks such as storytelling.
Moreover, these approaches cannot model how high-level narrative structures -- such as topics and text-visual alignment -- unfold over time in response to sequential visual input. Therefore, there is a need for automated methods to capture the temporal dynamics of narrative content and their alignment with evolving visual stimuli.

This work investigated specific cognitive and linguistic challenges by analyzing topical shifts, their temporal dynamics, and the consistency between narratives and corresponding visual stimuli.
We proposed two novel approaches to macrostructural narrative analysis: Dynamic Topic Models (DTM) for temporal analysis of topical evolution; and Text-Image Temporal Alignment Model (TITAN) for cross-modal (i.e., stimulus images and spoken narratives) alignment analysis.
These approaches aim to reveal cognitive deficits by identifying narrative impairments associated with NCDs, thereby enhancing our understanding of their impact on communication and cognition.
Our analysis was conducted primarily on the Cantonese CU-MARVEL-RABBIT corpus, with the English ADReSS~\cite{luz2020alzheimer} and ADReSSo~\cite{luz2021detecting} corpora included for cross-comparison with existing studies.
Experimental results on the CU-MARVEL-RABBIT corpus demonstrated the effectiveness of dynamic topic consistency as a macrostructure metric (F1=0.65, AUC=0.78).
The proposed TITAN model achieved superior performance (F1=0.72, AUC=0.81) in classification tasks, outperforming models based on handcrafted microstructure and macrostructure feature sets.
Moreover, comparative experiments further demonstrated the effectiveness of the proposed dynamic macrostructure modeling approaches for NCD detection (F1=0.8889 for ADReSS and F1=0.8504 for ADReSSo).

The paper is organized as follows: Section~\ref{sec:review} reviews related works; Section~\ref{sec:data} introduces the corpora used in this study; Section~\ref{sec:approach} illustrates the proposed approaches; Section~\ref{sec:exp} presents our experimental settings and results; Section~\ref{sec:discuss} analyzes and discusses experimental results; and Section~\ref{sec:conclusion} concludes this paper and outlines directions for future works.

\section{Related Work}
\label{sec:review}

Extensive research has been conducted in the domain of narrative-based NCD detection, leveraging various methodologies to identify biomarkers and assess cognitive impairments.
This section reviews pertinent literature encompassing both microstructural and macrostructural narrative analyses, collectively laying the groundwork for more comprehensive narrative assessment in NCD detection.
The term ``microstructure'' refers to fine-grained elements of narrative production, including lexicon, syntax, semantics, phonetics, acoustics, and other smaller linguistic units.
In contrast, the term ``macrostructure'' refers to higher-level aspects of narrative construction, such as global coherence, thematic development, and the overall organization of narratives.
Both microstructural and macrostructural analyses are crucial for a comprehensive understanding of cognitive deficits reflected in NCD narratives. 
However, existing studies are limited in their ability to capture the full spectrum of narrative features, particularly the dynamic aspects of narrative construction.
By identifying these limitations, this review highlights the gaps that the proposed methods aim to fill -- particularly through the incorporation of temporal dynamics and macrostructural consistency to advance the study of NCD narratives.

\subsection{Microstructural Analysis in NCD Detection}
 
Individuals with NCDs exhibit speech deficits that affect spoken language across both acoustic and linguistic dimensions. Prior research has explored the various aspects of these impairments in NCD speech, encompassing prosody (e.g., increased dysfluencies and longer, more frequent pauses), phonetics (e.g., abnormal Mel-Frequency Cepstral Coefficients patterns), phonology (e.g., higher rates of word mispronunciations), lexicon (e.g., reduced type-token ratio and elevated stopword ratio), syntax (e.g., reduced parse-tree depth), and semantics (e.g., abnormal patterns in word- and sentence-embeddings)~\cite{li2021comparative,adhikari2021comparative,kurdi2024dementia, cera2023speech,kang2024not, melistas23_interspeech, parsapoor_parsa_performance_2023, liu2023learning,pl2024cognitive,varlokosta2024methodologies,meng2023integrated,li2023leveraging, ivanova2023defying,kaltsa2024language,williams2023lexical,wang2024cross}. 
These microstructural features are essential for understanding the basic language abilities of individuals with NCDs and provide a foundation for narrative analyses in this domain.

Some recent efforts have focused on identifying highly correlated and significant biomarkers of NCDs from the aforementioned features~\cite{fraser2016linguistic,weiner2019speech,pulido2020alzheimer,frankenberg2021verbal,cho2022lexical}.
For example, Cho~\textit{et al.}~\cite{cho2022lexical} linked higher CSF p-tau\footnote{CSF p-tau: Cerebrospinal fluid phosphorylated tau, a NCD-related biomarker indicating the burden of neurofibrillary tangle pathology.} levels (indicating greater severity of 
NCD pathology) to specific acoustic and linguistic features: lower ratios of prepositions and nouns, higher pause rates, and shorter speech segment durations.
Wang~\textit{et al.}~\cite{wang2024cross} proposed several features based on structured clinical rating scales, such as dysfluencies, repetition, content relevance, syntactic structure and irrelevant information.
Such microstructural analysis provides valuable insights into local features, but more comprehensive information can be offered through macrostructural approaches that examine broader narrative structure and contexts.
Hence, this work extends prior research by examining the macrostructure of VSNs, including topic evolution and cross-modal consistency.

\subsection{Macrostructural Analysis in NCD Detection}

While microstructural analysis highlights specific language production impairments associated with NCDs, macrostructural analysis is equally important as it reveals the integrity of higher-order cognitive processes such as attention, working memory, planning, and organization. This macrostructural perspective is essential for understanding how NCDs affect the construction of coherent and well-organized narratives~\cite{glosser1991patterns,ellis2016global,cahana2018working,kong2023spoken}.

Macrostructural analysis in the literature generally focuses on measures of cohesion (the use of linguistic devices for structural unity) and coherence (logical flow and overall consistency).
Individuals with NCDs tend to produce narratives characterized by wrong co-referencing, inconsistencies in maintaining the main topic, uncompleted or irrelevant information, and disordered events~\cite{obler1983language,ripich1988patterns,ellis1996coherence,ripich2000conversational,st2005lack}.

Traditional macrostructural features were typically extracted manually based on expert annotations, such as readability~\cite{heitz-etal-2024-influence}, relationship between a proposition and previous one or the global topic~\cite{lima2014alzheimer}, or connection between events, macro-propositions and main information units~\cite{de2019evaluation}.

Recent advances leverage machine learning and LLMs for automatic analyze macrostructure of VSNs.
Some studies have quantified global semantic coherence through cosine similarity between adjacent utterance embeddings~\cite{burke2023comparing, parsapoor_parsa_performance_2023}, while others have assessed pragmatic coherence via reference-based metrics such as BLEU and METEOR~\cite{li2024devising}.
More recent work has adopted LLM evaluators to automatically predict macro-descriptors~\cite{botelho2024macro}.

However, as mentioned in Section~\ref{sec:intro}, these methods are unsuitable for NCD spoken narratives: similarity scores are confounded by frequent repetitions, reference-based metrics (BLEU, METEOR) are inapplicable due to non-standardized references, and LLM-based evaluation fails to meet interpretability requirements.
Additionally, limited research has investigated the temporal dynamics evolution of topics and the cross-modal alignment between visual stimuli and speech in VSN tasks. 
These gaps underscore the need for more advanced approaches to evaluate NCD narratives from both macrostructural and temporal perspectives, capable of capturing the interplay between evolving topics and the integration of visual and spoken elements.

\section{Corpora}
\label{sec:data}

This study is primarily conducted based on the Cantonese CU-MARVEL-RABBIT (an acronym for \underline{C}ognitive Assessment \underline{U}sing \underline{M}achine Le\underline{A}rning Empowe\underline{R}ed \underline{V}oic\underline{E} Ana\underline{L}ysis -- Rabbit Story Part) Corpus, with the English ADReSS and ADReSSo Corpora~\cite{luz2020alzheimer, luz2021detecting} included for comparative purposes.
The CU-MARVEL-RABBIT corpus comprises narratives elicited through the Rabbit Storytelling Task -- a 15-frame cartoon narrative prompt depicting a girl and her dog searching for a rabbit while encountering various animals and adventures (similar to ``\textit{Frog, Where Are You?}''~\cite{cameron1999frog}, but with fewer frames and more vibrant visuals). Similarly, the ADReSS and ADReSSo corpora contain speech recordings from the Cookie Theft Picture Description Task, which is based on a single static image depicting two children stealing cookies while a mother is distracted washing dishes~\cite{goodglass1972assessment}.

\subsection{The ADReSS \& ADReSSo Corpora}

The ADReSS~\cite{luz2020alzheimer} and ADReSSo~\cite{luz2021detecting} datasets, developed by the University of Pittsburgh, contain spontaneous speech recordings from elderly English-speaking adults in the Cookie Theft Picture Description task~\cite{goodglass1972assessment}.
% (Fig.~\ref{fig:data}(a)).
This task aims to assess participants' ability to recognize the elements in the picture and produce a coherent narrative based on their spatial relationships and implied events.

The ADReSS corpus consists of audio recordings and associated transcripts with 156 participants, which is partitioned into balanced training (54 HCs and 54 NCDs) and testing (24 HCs and 24 NCDs) subsets.
In contrast, the ADReSSo corpus includes a larger number of samples (237 participants), but does not provide any textual transcripts. It is partitioned into a training subset with 79 HCs and 87 NCDs, and a test subset comprising 36 HCs and 35 NCDs.

\subsection{The CU-MARVEL-RABBIT-RABBIT Corpus}

The CU-MARVEL-RABBIT corpus was designed and collected by our research group in Hong Kong as part of the broader CU-MARVEL project, which includes multiple cognitive assessment tasks.
To evaluate the ability of individuals with NCDs to integrate sequential visual information and generate coherent narratives, we utilize the Rabbit Story task within the CU-MARVEL framework, referred to herein as the ``CU-MARVEL-RABBIT'' corpus.

This corpus contains narratives from Cantonese-speaking participants telling a story illustrated through 15 cartoon images about a girl finding a rabbit (similar to ``\textit{Frog, Where Are You?}''~\cite{cameron1999frog}).
% , as illustrated in Fig.~\ref{fig:data}(b).
These images generally depict 5 sequential scenes (or topics): (1) the discovery and subsequent loss of the rabbit (images 1-4), (2) an encounter with a squirrel and some ants (images 5-7), (3) a meeting with a goose (images 7-9), (4) a run-in with a cow (9-11), and (5) a reunion with the rabbit followed by a farewell wave to its family (images 12-15).
Note that overlapping images (e.g., images 7 and 9) indicate transitional events, such as the emergence of the goose and the cow.

These images are presented in a booklet, with each image displayed on a separate page.
To ensure uniformity and comparability, all participants first review all 15 images to familiarize themselves with the content. Then, they examine each image individually, describing its content before moving on to the next. This process continues sequentially until narratives for all 15 images have been completed.

Illustrative examples are provided in the Appendix.
The participant in the HC group (Fig.~\ref{fig:hc_example}) tends to capture most visual details (e.g., characters (\textit{girl, dog, rabbit, squirrel, ants, goose, cow}); objects (\textit{tree hollow, rock}); and settings (\textit{house, garden, forest})) and weaves them into a clear, logical storyline across the images (e.g., \textit{rabbit was missing → girl and dog searched in the forest → rabbit was found with its family}).
In contrast, the participant with NCDs (Fig.~\ref{fig:major_ncd_example}) often misses or misidentifies key details (e.g., \textit{squirrel, goose}) and exhibits difficulties in establishing coherent connections between events, resulting primarily in isolated descriptions.

The CU-MARVEL-RABBIT Corpus (version ``2024-06-13'') consists of 758 participants aged over 60 years, including 467 HCs (mean age $\pm$ standard deviation: 71$\pm$7 years) and 291 NCDs (76$\pm$8 years).
It's partitioned into a training subset containing 427 HCs and 231 NCDs, and a test subset comprising 40 HCs and 60 NCDs.
To benchmark the performance of front-end processing tools (including speech recognition, word segmentation, and part-of-speech tagging), we extensively annotated a subset of 46 participants (22 HCs and 24 NCDs) from the test set (aligned with~\cite{meng2023integrated}), referred to as the ``CMR-Frontend-Evaluation'' set.

The CU-MARVEL-RABBIT Corpus also records the specialist neurological assessment for each participant, using a 0-4 severity scale for cognitive impairments, with lower values corresponding to better cognitive performance. Specifically, values 0-1 indicate HCs, while 2, 3, and 4 represent progressively severe NCDs.
Detailed data distribution and train-test split are listed in Table~\ref{tab:data_distribution}.

\begin{table}[htb]
    \centering
    \renewcommand{\arraystretch}{1.05} %
    \caption{Distribution of CU-MARVEL-Rabbit Corpus}
    \label{tab:data_distribution}
    \begin{tabular}{llcccccc}
    \toprule
    \multirow{2}{*}{\textbf{Split}} & \multirow{2}{*}{\textbf{Gender}} & \multicolumn{5}{c}{\textbf{Label}} & \multirow{2}{*}{\textbf{Total}} \\
    \cmidrule(lr){3-7}
    & & \textbf{0} & \textbf{1} & \textbf{2} & \textbf{3} & \textbf{4} & \\ 
    \midrule
    \multirow{3}{*}{\textbf{Train}} & \textbf{Male}   & 94 & 107 & 5  & 86 & 12 & \textit{304} \\ 
    & \textbf{Female} & 81 & 145 & 37 & 74  & 17  & \textit{354} \\ 
    & \textbf{Total}            & \textit{175} & \textit{252} & \textit{42} & \textit{160} & \textit{29} & \textbf{\textit{658}} \\ 
    \midrule
    \multirow{3}{*}{\textbf{Test}}  & \textbf{Male}   & 10  & 10  & 10 & 10  & 10  & \textit{50}  \\ 
    & \textbf{Female} & 10  & 10  & 10 & 10  & 10  & \textit{50}  \\ 
    & \textbf{Total}            & \textit{20}  & \textit{20}  & \textit{20} & \textit{20}  & \textit{20}  & \textbf{\textit{100}} \\ 
    \bottomrule
    \end{tabular}
    \vspace{-0.5em}
\end{table}

\section{Approach}
\label{sec:approach}
In this work, we propose a comprehensive framework for the automatic detection of NCDs using VSNs, encompassing both microstructural and macrostructural analyses. \textit{Microstructural analysis}, detailed in Section~\ref{sec:tradition}, focuses on the extraction of acoustic and linguistic features based on previous studies. In contrast, \textit{macrostructural analysis}, presented in Sections~\ref{sec:reference-based} through~\ref{sec:titan}, introduces three distinct methodologies aimed at capturing global and temporal narrative patterns.

\subsection{Traditional Acoustic and Linguistic Analysis}
\label{sec:tradition}

Microstructural features can be broadly categorized into two main groups: acoustic features and linguistic features.
Based on prior studies, we select 10 acoustic features and 13 linguistic features that are basic, universal and suitable for Cantonese in the CU-MARVEL-RABBIT Corpus, as summarized in Table~\ref{tab:biomarkers}.
Before extracting these features, we preprocess all speech recordings with FullSubNet+~\cite{chen2022fullsubnetp} denoiser and peak-normalization to ensure internal consistency and mitigate the risk of models learning spurious cues (i.e., the Clever Hans Effect~\cite{liu2024clever}).

Rows of A1-A10 in Table~\ref{tab:biomarkers} list the acoustic features, related to pause (A1-A6), speech duration (A7-A8), articulation rate (A9) and speaking rate (A10).
The articulation rate measures phonetic production speed as the number of syllables per voiced duration (excluding pauses), while the speaking rate quantifies overall delivery speed as the number of syllables per total duration (including pauses), thereby reflecting real-time communication fluency~\cite{jacewicz2009articulation}.

To extract these features, we first employ Voice Activity Detection (VAD) using the pretrained Feedforward Sequential Memory Network, as described in the paper~\cite{gao2023funasr}, to obtain the timestamps of the voiced segments.
With these segment timestamps, a pause is detected as the silence gap between two adjacent voiced segments when the duration exceeds the average syllable duration for that speaker, which is calculated by dividing the total voiced duration by the total number of syllables for each Cantonese participant.

The linguistic features are extracted from the automatic transcripts obtained using the pretrained Automatic Speech Recognition (ASR) model SenseVoice-Small~\cite{an2024funaudiollm}, which achieved a Word Error Rate (WER) of 31.37\% on the ADReSS test set and a Character Error Rate (CER) of 34.22\% on the CMR-Frontend-Evaluation set of the CU-MARVEL-RABBIT corpus.
Rows of L1-L13 in Table~\ref{tab:biomarkers} show the name and corresponding description of each feature.
To extract these linguistic features from the Cantonese transcripts, we used pretrained BERT-based models for word segmentation\footnote{\url{https://huggingface.co/ckiplab/bert-base-chinese-ws}} and PoS auto-tagging\footnote{\url{https://huggingface.co/ckiplab/bert-base-chinese-pos}}, achieving F1 scores of 82.69\% and 85.11\% for the CMR-Frontend-Evaluation set, respectively.

\begin{table}[tbh]
\centering
\caption{Microstructural features used in this work.\\
``\%'': ``ratio of'', ``\#'': ``number of'', ``\textnormal{dur}'': ``duration''.}
\label{tab:biomarkers}

\setlength\tabcolsep{1mm}
\resizebox{\linewidth}{!}{
\begin{tabular}{l l l}
\toprule
\textbf{ID} & \textbf{Feature Name} & \textbf{Description}                            \\
\toprule
\multicolumn{3}{c}{Acoustics} \\
\midrule
A1          & \textbf{\# pauses} & The number of pauses                                   \\
A2          & \textbf{Total pause dur} & all pause dur in seconds                   \\
A3          & \textbf{Average pause dur} & averaged pause dur in seconds            \\
A4          & \textbf{Normalized pause dur} & {\footnotesize Pause dur / articulation rate} \\
A5          & \textbf{Pause frequency} & \# pauses / speech dur                      \\
A6          & \textbf{Pause occurrence rate} & \# pauses / \# syllables             \\
A7          & \textbf{Total utterance dur} & all utterance dur in seconds                    \\
A8          & \textbf{Average utterance dur} & averaged utterance dur in seconds               \\
A9          & \textbf{Articulation rate} & \# syllables / voiced dur         \\
A10         & \textbf{Speaking rate} & \# syllables / total dur (include pauses)         \\
\toprule
\multicolumn{3}{c}{Linguistics} \\
\midrule
L1          & \textbf{\# words} & The number of words                     \\
L2          & \textbf{\% Stop words} & \# stop words / \# words                     \\
L3          & \textbf{\% Filled pauses} & \# filler words / \# words                \\
L4          & \textbf{\% Lexical filler} & \# lexical filler  / \# words            \\
L5          & \textbf{\% Backchannel} & \# backchannel words  / \# words            \\
L6          & \textbf{\% Repetition} & \# repeated words / \# words                 \\
L7         & \textbf{\% Adj} & \# adjectives / \# words                            \\
L8         & \textbf{\% Adv} & \# adverbs / \# words                               \\
L9          & \textbf{\% Noun} & \# nouns / \# words                               \\
L10          & \textbf{\% Pronoun} & \# pronouns / \# words                          \\
L11          & \textbf{\% Verb} & \# verb / \# words                                 \\
L12         & \textbf{\% Func} & \# functional words / \# words                               \\
L13         & \textbf{\% CTTR} & \# types / rooted(2 * \# tokens)     \\
\bottomrule
\end{tabular}
}
\vspace{-0.5em}
\end{table}

\subsection{Reference-based Narrative Alignment}
\label{sec:reference-based}
A critical part of evaluating the narrative task, particularly within the Rabbit Story context, involves assessing the comprehensiveness of a participant's narrative. To this end, we identify task-related keywords and measure their coverage in the narratives.
Specifically, we rank words from the HCs group (excluding punctuation and stopwords) based on their relevance to visual stimuli. This relevance is quantified using cosine similarity between global image embeddings (from the [CLS] token) and text embeddings (via mean pooling), both extracted from the final projection layers of the visual and textual encoders in the CLIP-like models (Chinese-CLIP\footnote{\url{https://huggingface.co/OFA-Sys/chinese-clip-vit-huge-patch14}}~\cite{yang2022chinese} for CU-MARVEL-RABBIT corpus, and Multilingual-CLIP\footnote{\url{https://huggingface.co/M-CLIP/XLM-Roberta-Large-Vit-B-16Plus}}~\cite{carlsson2022cross} for ADReSS and ADReSS corpora), which are pretrained via a contrastive learning paradigm that aligns images with corresponding textual descriptions.
As the textual and visual modalities are projected into a shared multimodal embedding space, the Chinese-CLIP model inherently enables cross-modal similarity measurement.
These derived top-ranked words are categorized into four semantic groups informed by human prior knowledge of the Rabbit Story, and the Narrative Scoring Scheme (NSS)~\cite{Helimann2009NSS}, including general settings or introduction (e.g., \textit{daytime, one day; garden, house, forest}), characters or roles (e.g., \textit{girl, dog, rabbit}), objects (e.g., \textit{sofa, box, stone}), and actions (e.g., \textit{play, find, walk}). Coverage (abbreviated as ``cvg'', ranging from 0 to 1) for each category is then computed as a feature for analysis, focusing on the narrative content rather than linguistic form (e.g., PoS ratios).

While these coverage metrics effectively capture whether local narrative elements are addressed, they provide limited insights into global coherence and connections across the entire picture set. To address this limitation, metrics commonly used in image captioning tasks, such as BLEU-n (n=1, 2, 3, 4), METEOR, and ROUGE-L, are adopted to quantify the degree of alignment between participants' narratives and reference descriptions.
For this purpose, we engaged six research team members, all thoroughly familiar with the NSS rubric\footnote{The NSS rubric evaluates narratives across seven dimensions: introduction, character development, mental and emotional states, referencing, conflict/resolution (or event/reaction), cohesion, and conclusion. Each dimension is rated on a 0-5 scale, with detailed elaborations and examples provided for the ``Proficient (5)'', ``Emerging (3)'', and ``Minimal/Immature (1)'' levels.}, to craft reference descriptions of the Rabbit Story that exemplify the proficient criteria.
The final BLEU-n, METEOR, and ROUGE-L scores are averaged across these references for each participant, providing additional features that reflect global narrative coherence.

In short, we derive two types of features for reference-based analysis:
(1) 10 domain-specific coverage metrics, including introduction, character, object, and action words, as well as high-frequency visual-related terms (including nouns, verbs, pronouns, adjectives, adverbs, and all PoS categories); and (2) 6 expert-aligned narrative similarity scores (BLEU-n, METEOR, ROUGE-L).
These reference-based features enable macrostructural analysis by evaluating informativeness and coherence against curated standards.
However, as these features derive primarily from word frequencies or local n-gram matching, they are incapable of modeling temporal topic development or the cross-modal consistency.

\subsection{DTM-based Temporal Analysis}
\label{sec:DTM-based}

\begin{figure*}[t]
    \centering
    \begin{minipage}[b]{0.24\linewidth}
    \centering
    \includegraphics[width=\linewidth]{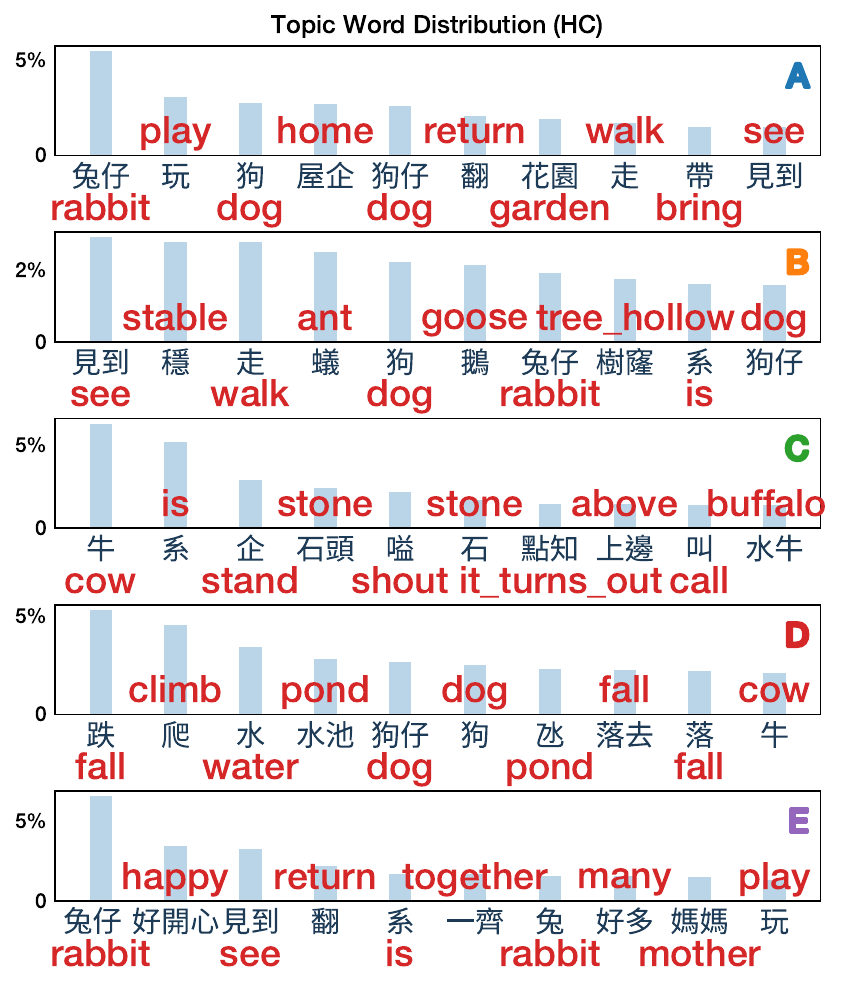}\\
    \vspace{-0.5em}
    {\small (a)}
    \end{minipage}
    \begin{minipage}[b]{0.24\linewidth}
    \centering
    \includegraphics[width=\linewidth]{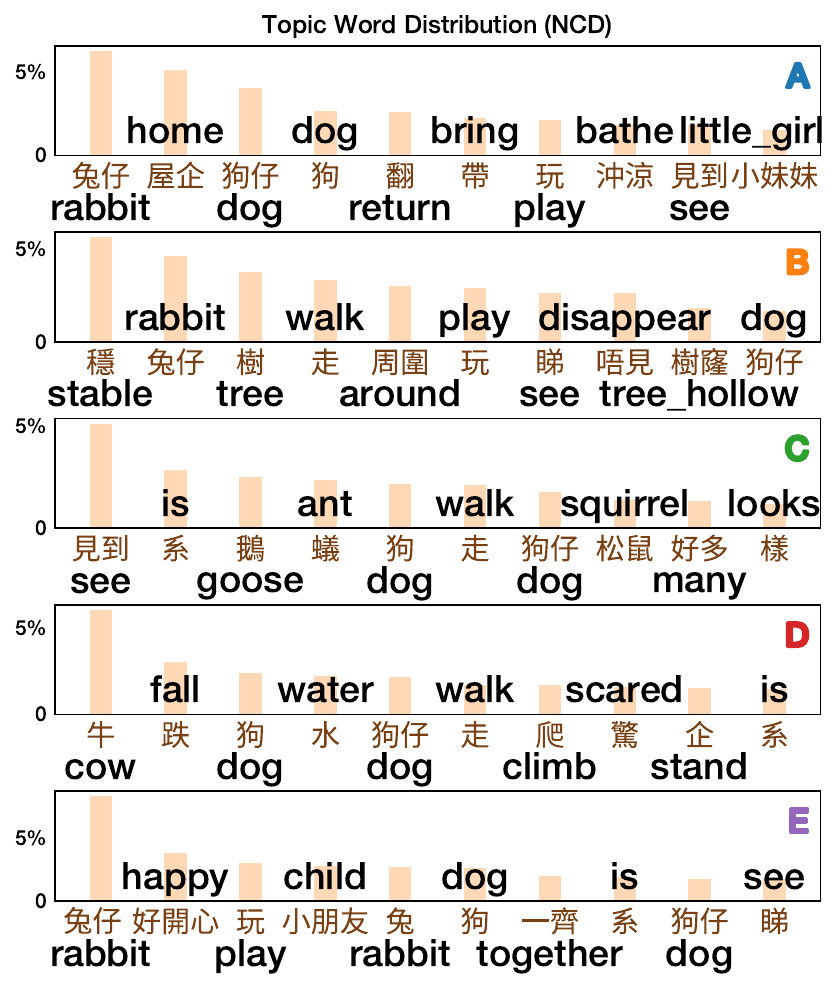}\\
    \vspace{-0.5em}
    {\small (b)}
    \end{minipage}
    \begin{minipage}[b]{0.24\linewidth}
    \centering
    \includegraphics[width=\linewidth]{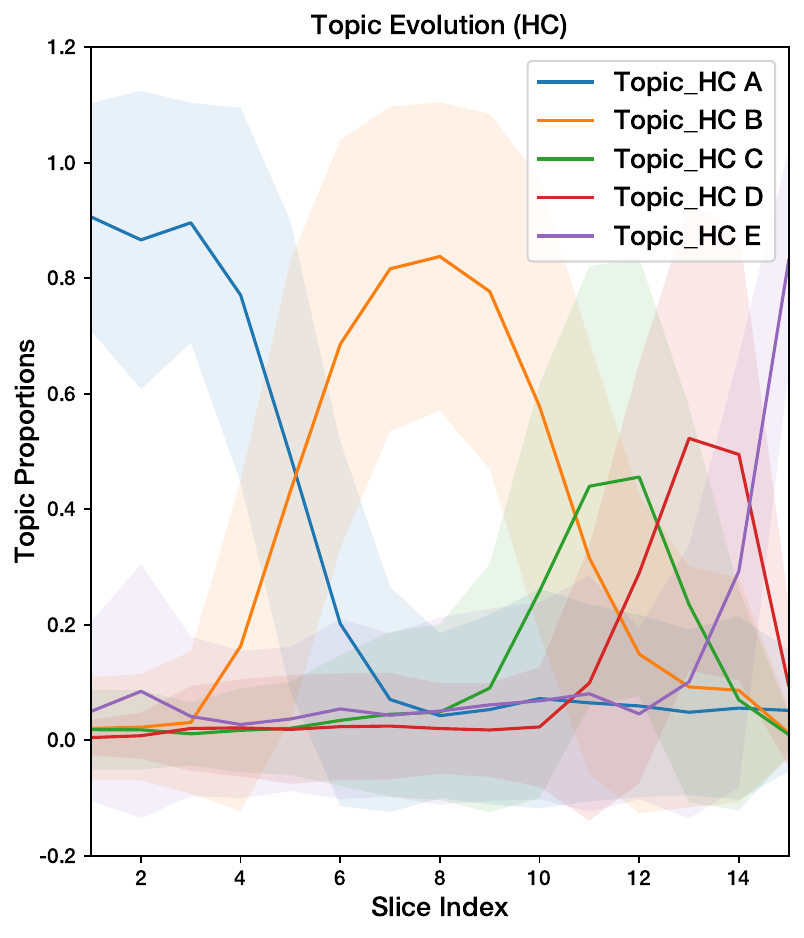}\\
    \vspace{-0.5em}
    {\small (c)}
    \end{minipage}
    \begin{minipage}[b]{0.24\linewidth}
    \centering
    \includegraphics[width=\linewidth]{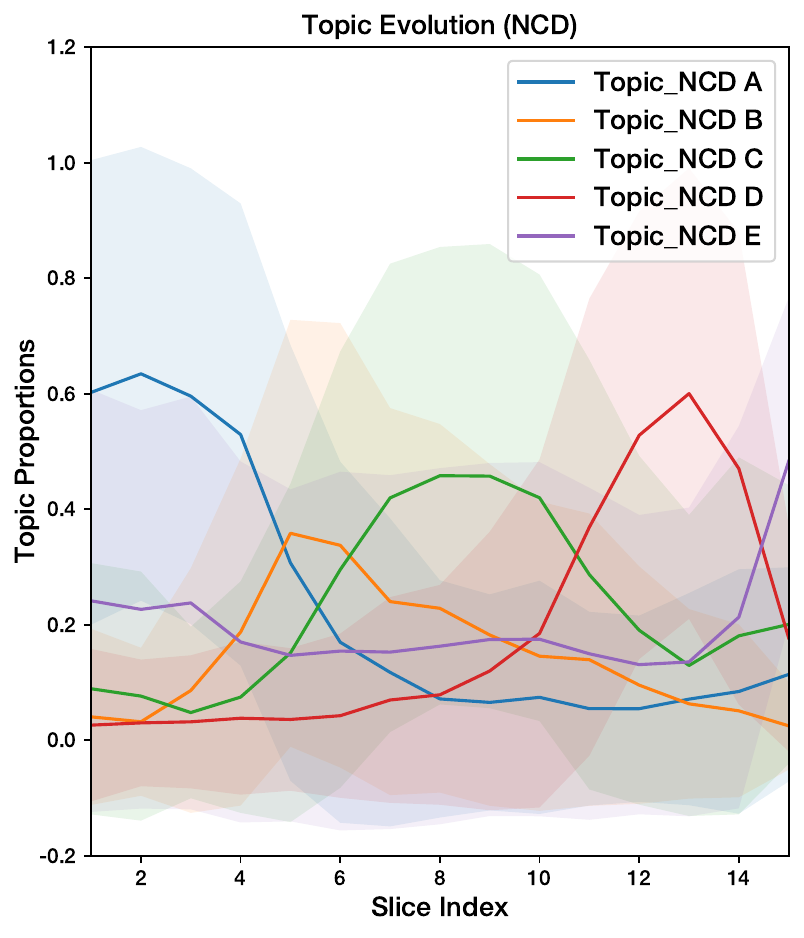}\\
    \vspace{-0.5em}
    {\small (d)}
    \end{minipage}
    \caption{The topic distribution for (a) the HC group and (b) the NCD group; and the topic evolution for (c) the HC group and (d) the NCD group; as obtained from DTM. Plots (a) and (b) show the weights of the top-10 topic words for each topic (A to E) for the group. Plots (c) and (d) display the average topic proportions with curves and standard deviations with shaded areas. Ordering of topics A to E is based on the positions of their first peaks in topic evolutions.}~\label{fig:dtm_evolution}
    \vspace{-1em}
\end{figure*}

Temporal changes or topic evolution in narratives are critical for understanding the progression and manifestations of NCDs, as these elements reflect higher-level cognitive abilities.
To capture these patterns, we utilize Dynamic Topic Modeling (DTM)~\cite{blei2006dynamic}, which tracks them by explicitly modeling the evolution of topics over time.
In contrast to previously mentioned static methods, DTM offers a dynamic perspective on the underlying temporal trends within the narratives.

To understand how DTM captures these temporal dynamics, it is important to first grasp the basics of topic modeling.
Topic modeling provides a probabilistic framework to uncover the latent semantic structure within textural data.
While the arbitrarily generated text may be randomly distributed across documents, the topic model assumes that words in natural languages are distributed around meaningful semantic topics. 
Specifically, document generation is assumed to be based on a probabilistic combination of underlying topics, where each topic is represented by a probabilistic distribution over vocabulary words. 
In the classical Latent Dirichlet Allocation (LDA), these distributions are assumed to be static and drawn from Dirichlet priors.

Dynamic Topic Model (DTM) is an extension of classical topic models. It introduces time dependencies upon the distributions, making it particularly suitable for analyzing temporal changes in narrative content.
Formally, DTM models the dynamic nature of topic evolution by introducing the time dependency $t$ in the topic word distribution $\beta_{k,t}$, where $k$ and $t$ denote the topic and time indices respectively.
In this way, it allows the topic distributions to drift over time via a Gaussian process, i.e.,
$\beta_t | \beta_{t-1} \sim \mathcal{N}(\beta_{t-1}, \boldsymbol{\Sigma})$,
where $\boldsymbol{\Sigma}$ represents the covariance matrix controlling topic transition smoothness.

The generative process of DTM proceeds as follows:

First, for each document $d$ in time slice $t$, the model first samples a topic proportion from a Dirichlet prior:
$\theta_d \sim \text{Dir}(\alpha)$,
where $\alpha$ is the Dirichlet concentration parameter controlling the sparsity of topic distributions.

Second, for each word position $n$ in document $d$, the model samples a topic assignment from a multinomial distribution parameterized by the document topic proportion:
$z_{d,n} \sim \text{Mult}(\theta_d)$.

Finally, the word is sampled from a multinomial distribution defined by the assigned topic's word distribution at that specific time slice:
$w_{d,n} \sim \text{Mult}(\beta_{t,z_{d,n}})$.

The model parameters are estimated through variational inference~\cite{blei2017variational} or the Expectation-Maximization~\cite{moon1996expectation} algorithm during the training phase.

In order to apply the DTM model to narrative analysis for NCD detection, we split each participant's ASR transcription evenly into 15 slices, assuming that each image leads to approximately the same number of words in spoken narratives.
To mitigate fragmented sentences, we shift the split points to the nearest punctuation boundaries.
This segmentation allows us to track the evolution of participants' narrative topics across the story's progression, enabling a granular analysis of discourse cohesion by observing how thematic content shifts and develops with each sequential image in the narrative.

We apply DTM by inputting the segmented transcription data, where each slice represents a distinct portion of the narrative associated with a time step. The DTM processes this data to identify evolving topics within and across the segmented slices, allowing us to observe how smoothly and coherently topics transition as the story progresses.

Fig.~\ref{fig:dtm_evolution} illustrates the topic distribution (a, b) and temporal evolution of topic proportions (c, d) for the HC and NCD groups.
For the distribution of topics over words, there exists differences between HC and NCD groups, where HC topics over time tend to be more aligned with the order of the images.
For example, the topic words of the HC group, shown in Fig.~\ref{fig:dtm_evolution}(a), include ``\textit{home}" and ``\textit{garden}" in topic ``A'' (corresponding to settings of images 1-4), ``\textit{ant}'', ``\textit{tree hollow}'' and ``\textit{goose}'' in topic ``B'' (corresponding to images 5-8), ``\textit{cow}'' and ``\textit{stone}'' in topic ``C'' (corresponding to images 9-10), ``\textit{fall}'' and ``\textit{pond}'' in topic ``D'' (corresponding to images 10-11), and ``\textit{many}'' (rabbits) and ``\textit{mother}'' in topic ``E'' (corresponding to images 12-15).
These topic words generally align with the different scenes along the story.
In contrast, the topic words of the NCD group, shown in Fig.~\ref{fig:dtm_evolution}(b), are more chaotic. For example, ``\textit{tree hollow}'' in topic ``B'' and ``\textit{ant, squirrel}'' in topic ``C'' both predominantly appear in the same scene in images 5 and 6, indicating overlapping topic assignments.

Another interesting finding is that the HC narratives exhibit a cyclical pattern.
Specifically, most HC participants mention ``\textit{return}'' and ``\textit{home}'' in topic "E", even though these elements are not depicted in the final images.
This suggests that HCs can tie the ending of the story back to its beginning, completing the girl's quest to find the lost rabbit and bringing the narrative to a satisfying closure.

For the dynamic proportions of these topics over time, as shown in Fig.~\ref{fig:dtm_evolution}(c, d), we observe that the HC group exhibits more distinct peaks (with each topic dominating consecutive slices) and smoother transitions between topics over time (peaks rising sharply after the preceding topic declines). In contrast, the NCD group shows more erratic changes with less pronounced peaks (e.g., reduced peak amplitudes and sporadic dominance).

\begin{figure*}[tbp]
    \centering
    \includegraphics[width=0.95\linewidth]{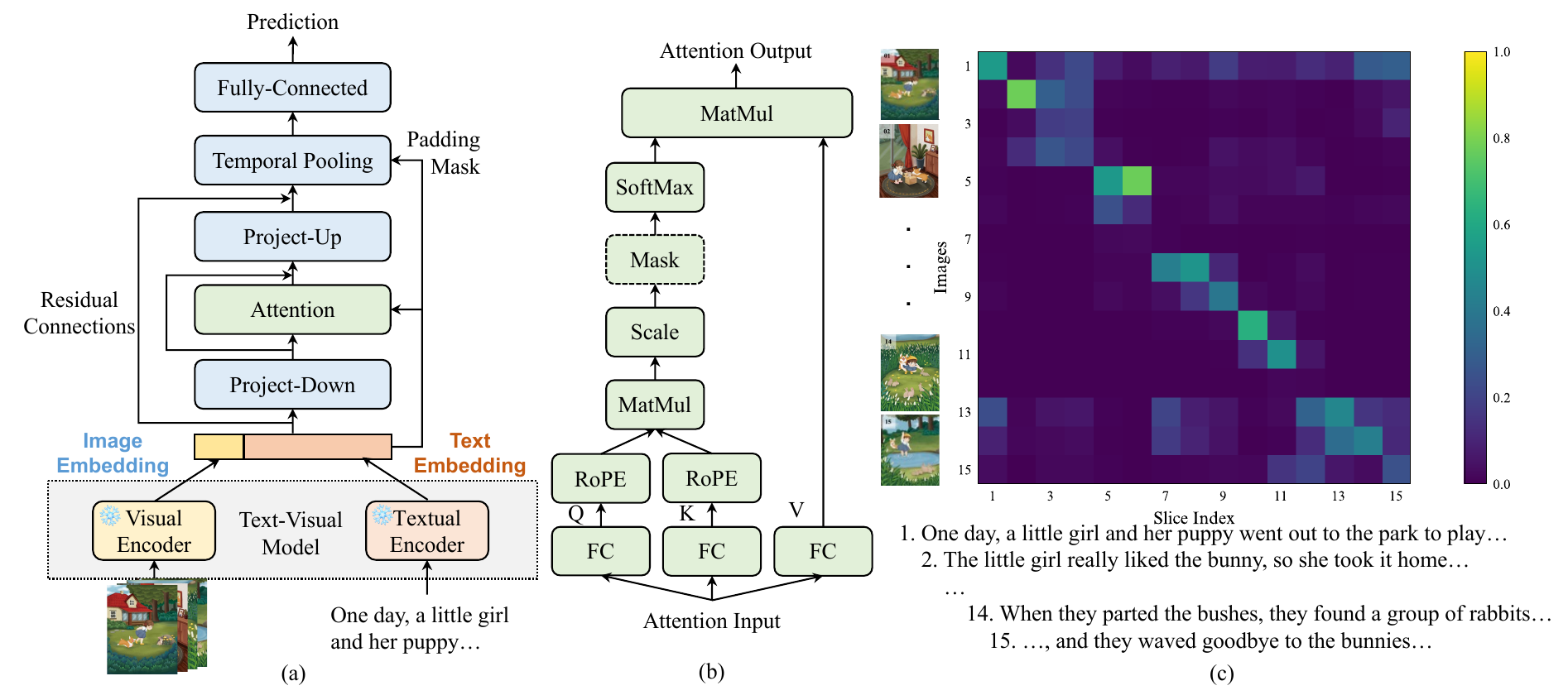}
    \vspace{-0.5em}
    \caption{(a) The proposed TITAN model. (b) The Attention module of TITAN. (c) The correlation map between image and text embeddings in TITAN, where the 15 textual chunks are manually segmented and aligned with the corresponding images. TITAN takes images and text narratives as inputs to predict NCDs, with the Attention mechanism using RoPE to emphasize important positions.}~\label{fig:framework}
    \vspace{-1em}
\end{figure*}

Based on these findings, we propose six DTM-related statistics for NCD detection.
The first metric, termed ``\textit{topic consistency}'', is derived from the hit rate of the top-10 topic words at each time step, derived from a DTM pretrained on all narratives.
This statistic measures how consistently a participant's topic words align with the top-10 words over time, thus providing insights into the stability and coherence of topics across different groups (HC vs. NCD).
For example, if the words in the text slice include all top-10 topic words at the same timestamp $t$, then topic\_consistency($t$) = 1.0.
Higher topic consistency suggests a more stable thematic structure, indicating a potentially healthier cognitive state.
The second metric, termed ``\textit{topic cycle}'', is a Boolean indicator of whether the word ``home'' appears in the latter half of each narrative.

The remaining four metrics capture the dynamic properties of topic evolution: (1) the standard deviation of topic proportions (\textit{topic variability}); (2) the temporal correlation of topic proportions over time (\textit{topic temporal corr}); (3) the peak-to-peak range, defined as the difference between the maximum and minimum values (\textit{topic ptp range}); and (4) the average rate of change, measured as the Euclidean distance between adjacent topic distributions over time (\textit{topic change rate}).
These statistics capture reflect the development of VSN topics, and can be used as features followed by a downstream classifier for NCD detection, while remaining interpretable through their well-defined mathematical properties.

\subsection{TITAN: Text-Image Temporal Alignment Network}
\label{sec:titan}

While DTM demonstrates potential in capturing topic evolution patterns, it faces challenges in incorporating cross-modal information in tasks such as picture description and VSNs.  
This limitation impairs topic representation modeling, preventing the effective utilization of image information and the handling of cross-modality correspondence.  
Previous studies have attempted to address these issues by inputting images into multimodal large language models to generate reference descriptions and keywords~\cite{li2024devising}. However, these approaches often overlook temporal dynamics and fail to capture topic evolution throughout the entire narrative.  
To overcome these challenges, and drawing inspiration from multimodal metrics like CLIPScore and GROOViST~\cite{hessel2021clipscore, surikuchi2023groovist} (which measure the text-image alignment through contrastive learning), we propose the Text-Image Temporal Alignment Network (TITAN). TITAN treats images as dynamic topics and constructs temporal relationships using the Attention mechanism, as illustrated in Fig.~\ref{fig:framework}.

This network comprises several key components: (1) feature extraction, (2) embedding fusion, (3) projection layers, (4) Attention module, and (5) prediction head.
We describe each component in detail below.

\customsection{D1. Feature Extraction}

The first step involves extracting features from the input text and images using the pretrained text-visual models, comprising textual-encoder and visual-encoder to process different modalities. In this work, we adopted the ViT visual-encoder and RoBERTa textual-encoder from the pretrained CLIP-like models, as mentioned in Section~\ref{sec:reference-based}.
The CLIP-like models can effectively align the embeddings of images and corresponding texts in the semantic space, thus providing insights for narrative coherence (structural dynamics) and thematic consistency (content alignment).
Specifically, given an input image $I$ and text $T$, we obtain the embeddings $E_I$ (from the [CLS] token) and $E_T$ (via mean pooling) from the final projection layer of their respective encoders:

\vspace{-0.5em}
$$
E_I = \text{CLIP}_\text{Visual}(I) \in \mathbb{R}^{J\times H}, E_T = \text{CLIP}_\text{Textual}(T) \in \mathbb{R}^{K \times H}
$$
\vspace{-1em}

where $H$ is the hidden dimension of the extracted embeddings, $J$ is the number of image frames (e.g., $J=1$ for the ADReSS and ADReSSo corpora and $J=15$ for the CU-MARVEL-RABBIT corpus), and $K$ is the temporal length of the investigated text sequence.

\customsection{D2. Embedding Fusion}

The extracted embeddings $E_I$ and $E_T$ are then concatenated to form a combined representation:

$$
E_{IT} = [E_I; E_T] \in \mathbb{R}^{(J+K)\times H}
$$

where $[;]$ denotes the concatenation operation.
This concatenation operation offers at least two advantages. First, it allows the subsequent layers to share the same projection layers for both image and textual embeddings, which is parameter-efficient and feasible since the embeddings are aligned to a common multimodal space during the pre-training stage. Second, inputting the concatenated embeddings into the Attention mechanism enables the model to capture combined information from image-image, image-text, and text-text pairs, thereby enhancing the richness of the contextual features.

\customsection{D3. Projection Layers}

The combined embedding $E_{IT}$ is dimensionally adjusted through multiple projection layers, starting with a linear projection to reduce the embedding size:
$$
E_{down} = E_{IT}W_{down}^T + b_{down} \in \mathbb{R}^{(J+K)\times H^{\prime}}
$$
where $H^{\prime}$ is the reduced dimension, $W_{down} \in \mathbb{R}^{H^{\prime}\times H}$ and $b_{down} \in \mathbb{R}^{H^{\prime}}$ are the weight matrix and bias of the projection layer, respectively.

After the Attention module, the embedding is then up-projected to its original dimension for residual connections:
$$
E_{up} = E_{attn}W_{up}^{T} + b_{up} \in \mathbb{R}^{(J+K)\times H}
$$
where $W_{up} \in \mathbb{R}^{H\times H^{\prime}}$ and $b_{up} \in \mathbb{R}^{H}$ are the parameters of the up-projection layer. 

This bottleneck projection structure effectively balances dimensional reduction for computational efficiency and dimensional expansion for capturing complex feature interactions.

\customsection{D4. Attention Module}

As shown in Fig.~\ref{fig:framework}(b), the Attention module aims to capture the temporal dynamics and contextual relationships within and between modalities, by attending to specific positions of the projected embeddings.

To effectively process dynamic temporal information, we incorporate Rotary Position Embedding (RoPE)~\cite{su2024roformer} into the Attention mechanism. RoPE encodes positional information into token embeddings without disrupting the invariance properties of Self-Attention, making it well-suited for capturing temporal dependencies.
Mathematically, the positional encoding vector for the $i$-th position, $p_i \in \mathbb{R}^{2H^{\prime}}$, is defined using sinusoidal functions:
$$
p_i = [\sin(i), \cos(i), \ldots, \sin(\frac{2H^{\prime}}{2}i), \cos(\frac{2H^{\prime}}{2}i)]^{T}
$$

Each element of $p_i$ is derived from sinusoidal functions to capture positional information: $\sin(i)$ and $\cos(i)$ are applied to alternate dimensions, with increasing frequencies across dimensions to capture information at various scales.
Given the different temporal strides of images and texts, RoPE is applied separately to each modality with different frequencies, i.e., $p_i \in \mathbb{R}^{1, \ldots, H^{\prime}}$ for images and $p_i \in \mathbb{R}^{H^{\prime}+1, \ldots, 2H^{\prime}}$ for texts.

Given the projected embedding $E_{down}$, we compute the queries, keys, and values as follows:
$$
Q = E_{down}W_Q^T, K = E_{down}W_K^T, V = E_{down}W_V^T
$$
where $W_Q, W_K, W_V \in \mathbb{R}^{H^{\prime} \times H^{\prime}}$ are learnable parameter matrices for query, key, and value projections respectively.
The queries and keys, for image and text parts separately, are then transformed using RoPE to compute position-aware Attention scores.
Specifically, for each query $q_i$ and key $k_i$, the transformed versions are computed as:
$$
q_i^{\prime} = q_i \cos(p_i) + q_{i}^{\perp} \sin(p_i), k_i^{\prime} = k_i \cos(p_i) + k_{i}^{\perp} \sin(p_i)
$$
where $q_{i}^{\perp}$ and $k_{i}^{\perp}$ represent the perpendicular components of the query and key vectors.

Finally, the dot-product Attention is computed as follows:
$$
E_{attn} = \text{softmax}  \left( \text{Mask}(Q^{\prime} K^{\prime T}) / \sqrt{H^{\prime}} \right) V
$$
where $Q^{\prime}$ and $K^{\prime}$ are the RoPE-transformed queries and keys. The mask ensures that the padding positions of input get near-zero Attention weights after the softmax.
The scaling factor $\sqrt{H^{\prime}}$ prevents the dot products from growing too large in magnitude.
The values remain untransformed to preserve the original contents that are being passed through the network.

In VSN tasks such as the Rabbit Story, narrative transcripts must align with reference images. Participants with NCDs may misidentify animal names (anomia) or omit detailed scenarios (visual perception deficts) in these images. For example, many NCDs misidentify ``goose'' as ``duck'', or skip the event of ``goose in the brush'' in the seventh image.
The Attention module in the proposed text-image temporal alignment network is able to perceive potential misalignment and discrepancies cross modalities, and therefore recognize NCD participants with the consideration of their narrative macrostructure.

\customsection{D5. Prediction Head}

Finally, the up-projected embedding is temporally pooled with the Attention mask ($\in \mathbb{R}^{J+K}$) and passed through a fully connected layer to generate the prediction:
$$
E_{pool} =  \frac{\sum_{i=1}^{J+K} \text{mask}_i \cdot E_{up, i}}{\sum_{i=1}^{J+K} \text{mask}_i}
$$
$$
\hat{y} = \sigma(E_{pool}W_{fc}^{T} + b_{fc}) \in \mathbb{R}^{C}
$$
where $C$ is the prediction dimension (number of classes for classification tasks, or 1 for the regression task), $W_{fc} \in \mathbb{R}^{C\times H}$ and $b_{fc}\in \mathbb{R}^{C}$ are the weight and bias of the fully connected layer, and $\sigma$ is an activation function (softmax for classification tasks and identity for the regression task).

Throughout the network, residual connections are applied at each stage to facilitate gradient flow and mitigate the vanishing gradient problem.

The TITAN model leverages high-quality cross-modal representations from CLIP-like models as input features, which inherently capture strong image-text correlations. As shown in Fig.~\ref{fig:framework}(c), the heatmap visualizes the pairwise similarity between text chunks (x-axis) and image sequences (y-axis) computed from the raw CLIP embeddings $E_T$ and $E_I$.
The diagonal intensity pattern indicates that these input features inherently exhibit temporal and cross-modal consistency in narrative structures.
Bright spots along the diagonal reflect strong alignment between images and corresponding text segments, suggesting that events or scenes unfold in a logically ordered sequence.
These initial representations are further refined through the joint design of the projection, Attention, and prediction modules, enabling TITAN to capture cross-modal consistency and temporal coherence for NCD detection.

\section{Experiments \& Results}
\label{sec:exp}

\begin{table*}[tb]
\centering
\renewcommand{\arraystretch}{1.05} %
\caption{Classification and Regression Results on the CU-MARVEL-RABBIT corpus. Feature dimensions are given in parentheses.\\``*'' denotes baseline systems; ``$\downarrow$'' indicates that the lower values are better; ``\textnormal{Emb.}'' refers to embedding(s).\\The best results are highlighted in \textcolor{red}{red}, and the second-best are \underline{underlined}.}~\label{tab:classification_results}
\vspace{-1.5em}
\begin{tabular}{clcccccc|cc}
\toprule
\multirow{2}{*}{\textbf{Sys.}} & \multirow{2}{*}{\textbf{Features}} & \multirow{2}{*}{\textbf{Model}} & \multicolumn{5}{c}{\textbf{Classification}} & \multicolumn{2}{c}{\textbf{Regression}} \\
\cmidrule(lr){4-8} \cmidrule(lr){9-10}
 & & & \textbf{F1} & \textbf{AUC} & \textbf{Rec.} & \textbf{Prec.} & \textbf{Acc.} & \textbf{R2} & \textbf{RMSE$\downarrow$} \\
\midrule
1*  & Acoustics (10)       & SVM       & 0.6000 & \underline{0.8054} & 0.4500 & 0.9000 & 0.6400 & 0.1580  & 0.3244 \\ 
2*  & Linguistics (13)     & SVM       & 0.5376 & 0.6596 & 0.4167 & 0.7576 & 0.5700 & -0.0051 & 0.3544  \\ 
3*  & Micro (23)           & SVM       & 0.5843 & 0.7571 & 0.4333 & 0.8966 & 0.6300  & 0.1604 & 0.3240  \\ 
4  & Reference-based (16)  & SVM       & 0.6458 & 0.7946 & 0.5167 & 0.8611 & 0.6600  & 0.2112 & 0.3140  \\ 
5  & DTM-based (6)         & SVM       & 0.6522 & 0.7825 & 0.5000 & \underline{0.9375} & 0.6800  & 0.1952 & 0.3172  \\ 
6  & Macro (22)            & SVM       & \underline{0.6947} & 0.8029 & \underline{0.5500} & \textcolor{red}{0.9429} & \textcolor{red}{0.7100} & 0.2233 & \underline{0.3116}  \\ 
7  & All statistics (45)   & SVM    & 0.6875 & 0.7783 & \underline{0.5500} & 0.9167 & \underline{0.7000} & \underline{0.2367} & \textcolor{red}{0.3089}  \\
8 & Text-Visual Emb. (1024)     & TITAN & \textcolor{red}{0.7238} & \textcolor{red}{0.8120} & \textcolor{red}{0.6333} & 0.8444 & \textcolor{red}{0.7100}  & \textcolor{red}{0.3876} & 0.3130  \\ 
\bottomrule
\end{tabular}
\end{table*}

\subsection{Experimental Setup}
In this study, we conduct both classification and regression tasks to evaluate the performance of various feature sets and models. Eight systems are designed, each utilizing a distinct feature set. System 1 uses 10 acoustic features, System 2 incorporates 13 linguistic features, and System 3 combines these microstructural features. System 4 utilizes 16 proposed reference-based metrics, and System 5 employs six proposed DTM-based features. System 6 combines these macrostructural features, while System 7 integrates all microstructural and macrostructural features. As these features are all statistical in nature, we first apply Principal Component Analysis (PCA) with five components to reduce dimensionality and then utilize a Support Vector Machine (SVM) for prediction, with a Radial Basis Function (RBF) kernel and a soft margin $C=1$.
The number of principal components in PCA and the SVM parameters (namely, the ``kernel'' and ``soft margin'') are tuned via grid search using System 7 with all statistical features.
System 8, the TITAN model, as depicted in Fig.~\ref{fig:framework}, has a fully-connected linear layer as the task head.
To align with PCA in Systems 1 to 7, the projection dimension $H^{\prime}$ of TITAN is also set as five.

During training, the models are optimized for 100 epochs with an initial learning rate of 1e-3 and a weight decay of 1e-2.
The pretrained CLIP models are frozen, except the last projection layer of the textual encoder, which is fine-tuned with a learning rate of 1e-5. The batch size is set to 64 for the ADReSS and ADReSSo corpora and 256 for the CU-MARVEL-RABBIT corpus to accommodate differences in data availability and computational efficiency.

\subsection{Evaluation Protocol}
For evaluation, we employed multiple metrics across both classification and regression experiments.
For the classification task, the primary metrics were the F1-score and AUC, supplemented by recall (Rec.), precision (Prec.), and accuracy (Acc.), reported on the positive class (NCD). Except for AUC, the classification metrics were computed with a default threshold set at 0.5.
For regression, we normalized the ground-truth labels to the range [0, 1] and used the coefficient of determination ($R^2$, ranging from $-\infty$ to 1) and Root Mean Square Error (RMSE, ranging from 0 to $+\infty$) as metrics.
To ensure robustness in evaluation, results were averaged over the five final epochs of each experiment.
Notably, the testing sets of ADReSS, ADReSSo, and CU-MARVEL-RABBIT were held out from training and approximately balanced in terms of positive and negative samples, ensuring an unbiased evaluation.

\subsection{Results}

Table~\ref{tab:classification_results} presents the results of classification and regression tasks, comparing various feature sets and models using F1 as the primary evaluation metric. Among the systems, the proposed TITAN model (Sys. 8), which utilizes combined text and image embeddings, achieves the best overall performance in both tasks, with an F1 score of 0.7238 and AUC of 0.8120 in classification; and R2 of 0.3876 in regression.
In contrast, the linguistic feature-based system (Sys. 2) shows an inferior performance among these systems, which is limited by the ASR, segmentation and tagging components that are not tailored to colloquial Cantonese narratives.

Another key finding is that macrostructural statistics (Sys. 6) outperform microstructural statistics (Sys. 3), achieving F1 scores of 0.6947 versus 0.5843 in classification; and R2 of 0.2233 versus 0.1604 in regression. This highlights the effectiveness of discourse-level analysis in capturing NCD-related narrative characteristics.

\section{Analysis \& Discussion}
\label{sec:discuss}

In this section, we discuss the proposed DTM-based features and the detailed mechanisms of TITAN.
We first present an instance-level case study to illustrate these DTM-based features.
We then extend the analysis to all test examples to examine the correlation between the statistical features and NCDs, as well as their contributions to the classification model's performance.
Next, we investigate the interpretability of TITAN through the visualization of its Attention mechanism, and evaluate the impact of individual components via ablation studies.
Finally, we contextualize our findings by comparing TITAN with prior research on the ADReSS and ADReSSo corpora.

\begin{figure*}[ht]
    \centering
    \includegraphics[width=0.95\linewidth]{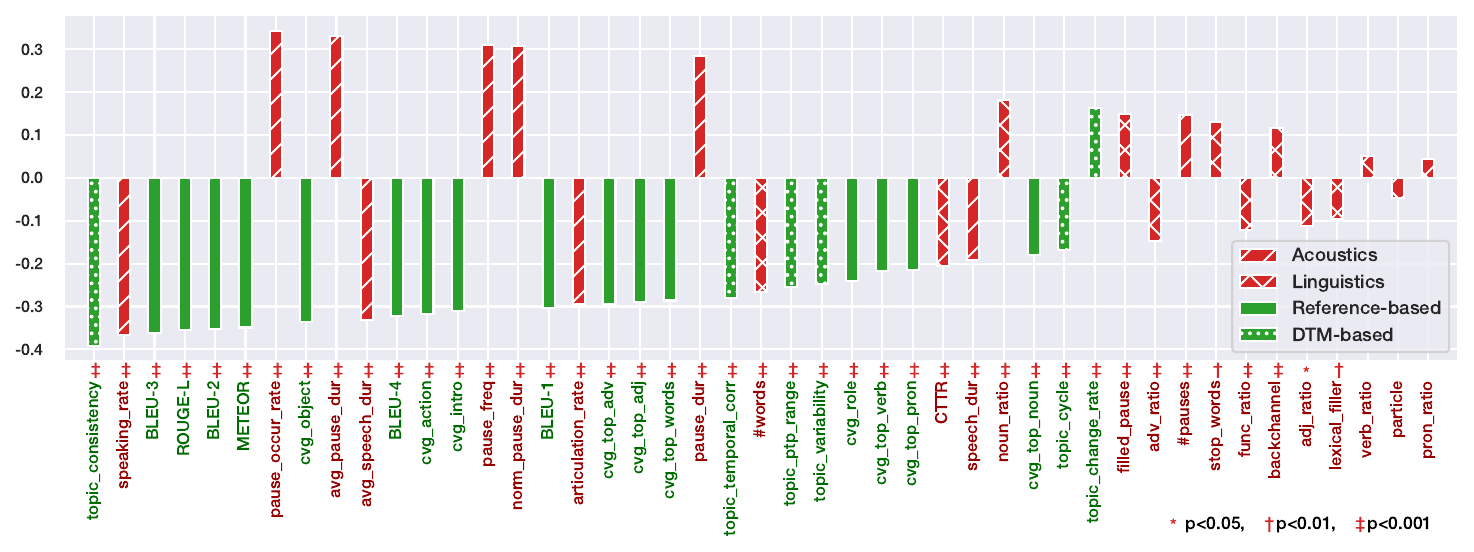}
    \vspace{-1.5em}
    \centering
    \caption{Correlation coefficients between proposed features and NCD labels, with statistical significance shown on the x-axis. The heights of the green (macrostructural) and red (microstructural) bars represent correlation values, indicating that most top-ranked features are macrostructural.}~\label{fig:feature_rank}
    \vspace{-1em}
\end{figure*}

\subsection{Case Study}

To facilitate an intuitive understanding of the DTM-based features introduced in Section~\ref{sec:DTM-based}, we present a comparative case study between a healthy control participant and a participant with NCDs from the CU-MARVEL-RABBIT corpus.
The transcripts and normalized feature values for these two examples are shown in Figs.~\ref{fig:hc_example} and~\ref{fig:major_ncd_example} in the Appendix.

As shown in Fig.~\ref{fig:hc_example}, the HC participant produced a coherent and lexically diverse narrative.
The transcript captures most salient visual elements, corresponding to a high ``\textit{topic consistency}'' score (0.50).
The mention of ``\textit{back home}'' is reflected in a positive ``\textit{topic cycle}'' value (normalized as 1.26).
High ``\textit{topic variability}'' (1.37) indicates rich lexical diversity within each topic distribution.
The ``\textit{topic ptp range}'' (1.12) reflects that specific topics are prominently expressed at each time step, i.e., topic distributions across time steps exhibit high peak-to-peak amplitudes.
The negative ``\textit{topic change rate}'' (-0.47) suggests smooth, gradual topic transitions.
The strong ``\textit{topic temporal correlation}'' (0.97) indicates that topic progression follows a coherent, time-ordered storyline.

In contrast, the NCD narrative (Fig.~\ref{fig:major_ncd_example}) omits or misidentifies key elements (e.g., incorrect animals such as cat, duck, chick, and sheep), resulting in low ``\textit{topic consistency}'' (-0.78).
The absence of narrative closure (``\textit{return home}'') leads to a negative ``\textit{topic cycle}'' value (-0.80).
Low ``\textit{topic variability}'' (-2.46) indicates limited vocabulary for each topic.
A high ``\textit{topic change rate}'' (0.56) and a negative ``\textit{temporal correlation}'' (-1.61) indicate incoherent topic development.

\subsection{Statistical Features Correlation Ranking}

Building upon the insights from the case study, we extend our analysis to assess how the proposed statistical features are associated with NCDs across all examples in the CU-MARVEL-RABBIT corpus.
To this end, we conduct Spearman's rank correlation test across features and ranked them by the absolute value of correlation coefficients. As illustrated in Fig.~\ref{fig:feature_rank}, the correlation distribution shows that the macrostructural features (green bars) exhibit greater importance compared to microstructural features (red bars). Notably, the DTM-based metric ``\textit{topic consistency}'', the acoustic feature ``\textit{speaking rate}'', and several reference-based coverage metrics ranked relatively high with strong statistical significance ($p<0.001$), suggesting their potential as indicators of NCDs.
Furthermore, many features associated with narrative macrostructural abilities display highly negative correlations with NCDs, whereas most pause-related features show positive correlations with NCDs. These findings highlight the utility of these features not only for effective classification, but also for gaining deeper insight into NCD-related narrative characteristics.

\subsection{Interpretable Feature Contribution Analysis}

To elucidate the role of the proposed features in the model's decision pathways, we quantify their contributions using SHapley Additive exPlanations (SHAP) values~\cite{Scott2017shap}.
SHAP is a model-agnostic explanation framework rooted in cooperative game theory, which fairly and consistently attributes a model's prediction to each input feature using Shapley values.
Specifically, for each feature $x_i \in X$ (the complete feature set), the SHAP value $\phi_i$ represents its average marginal contribution to the model's predictions when the target feature is added to all possible subsets $S$ of the other features, i.e.,
$$\phi_i = \sum_{S \subseteq X \backslash \{x_i\}} w_i {[f(S \cup \{x_i\}) - f(S)] },$$
where $w_i = |S|!(|X|-|S|-1)!/|X|!$ denotes the weight accounting for subset permutations, $|\cdot|$ denotes set cardinality, and $f(S)=ln{\frac{P(\hat{y}|S)}{1-P(\hat{y}|S)}}$ is the log-odds ratio of the model's output probability $P$ based on given inputs $S$.

\begin{figure}[ht]
    \centering
    \includegraphics[width=\linewidth]{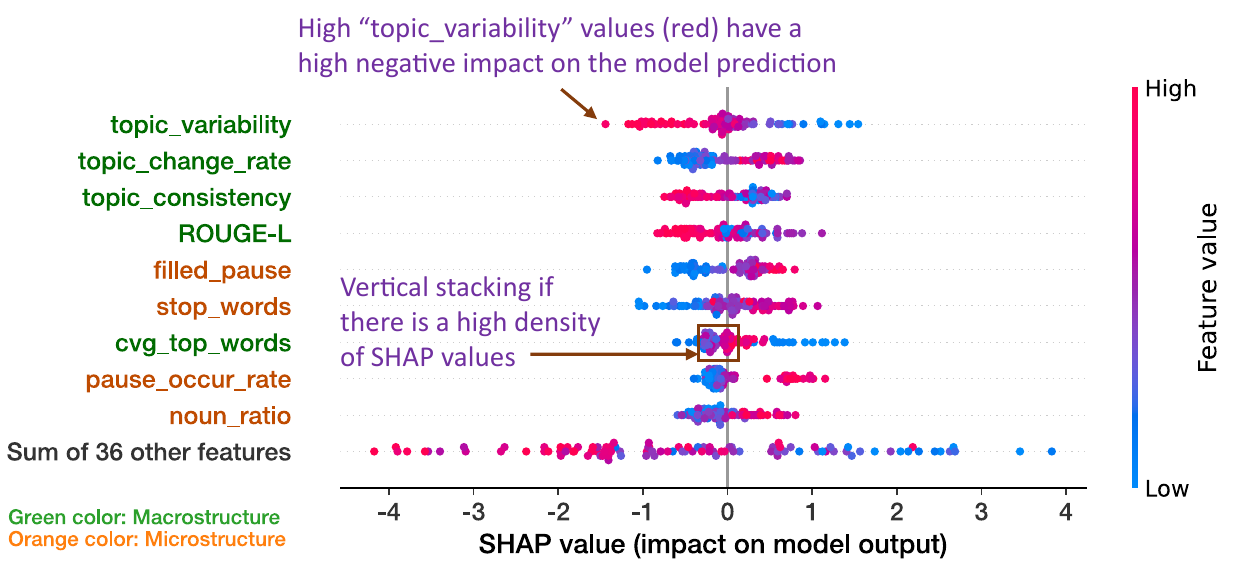}\\
    \vspace{-1em}
    \caption{Global SHAP value distribution (beeswarm plot) across all test samples in the CU-MARVEL-RABBIT corpus. Features are ranked vertically by mean absolute SHAP values (most impactful at the top). Each point represents an instance: the horizontal position indicates the SHAP value (negative or positive impact), the color reflects the standardized feature value (blue = low, red = high), and density of points along each feature indicates value distribution.}
    \label{fig:beeswarm}
    \vspace{-1em}
\end{figure}

This formulation enables global feature importance analysis by aggregating SHAP values over the held-out test set.
Fig.~\ref{fig:beeswarm} presents the global feature contributions for System 7, ranked by their absolute SHAP values.
Notably, the proposed macrostructural features, including ``\textit{topic variability}'', ``\textit{topic change rate}'', ``\textit{topic consistency}'' and ``\textit{ROUGE-L}'', emerged as the top-ranked global contributors.
These findings highlight the critical role of macrostructural patterns in distinguishing between individuals with NCDs and HCs.
The consistent directionality of their SHAP values (positive or negative associations with NCD classification) further validates their robust discriminative power across diverse instances.

\subsection{Attention Mechanism in Cross-modal Consistency}

\begin{figure}[ht]
    \centering
    \includegraphics[width=\linewidth]{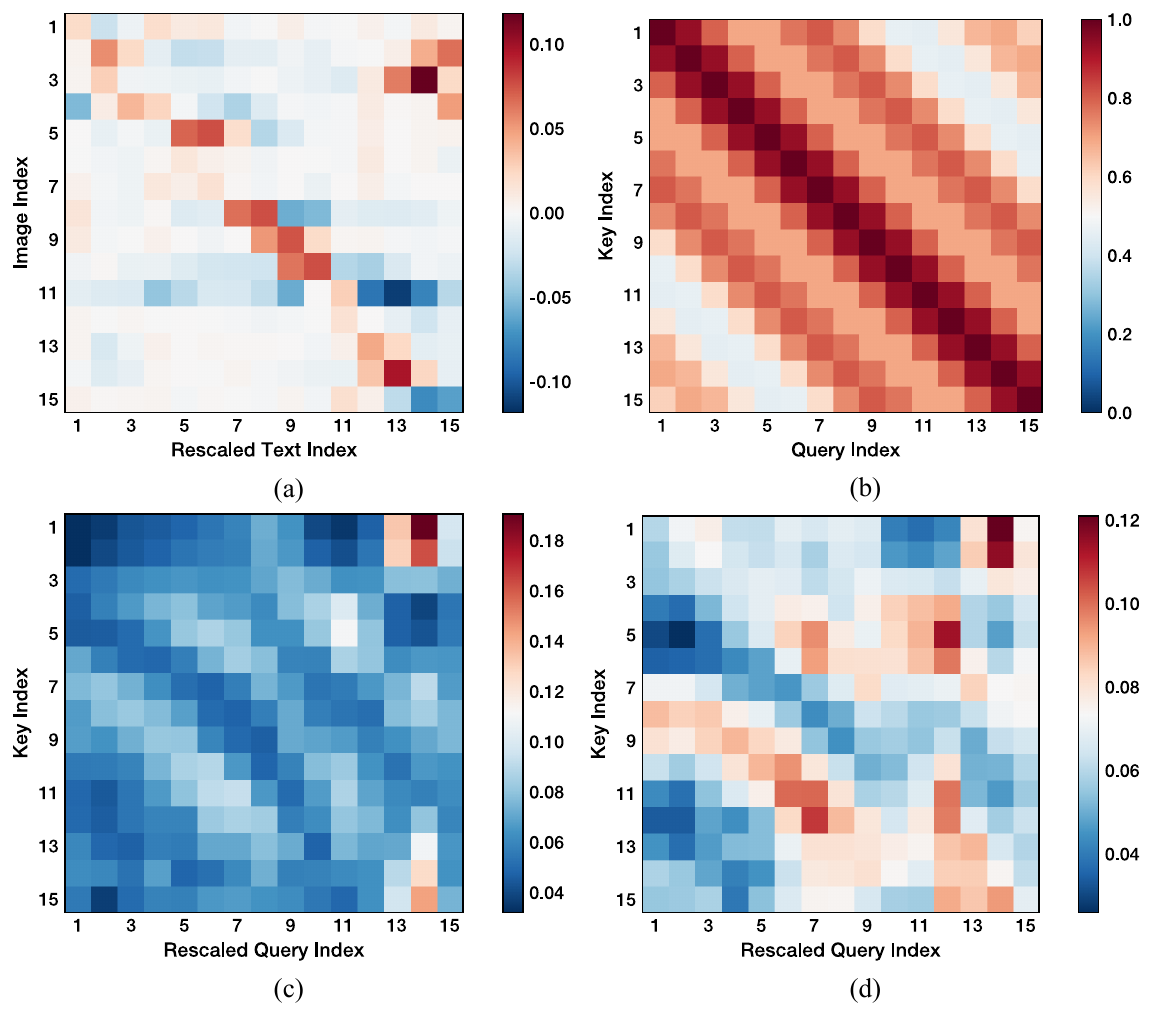}
    \vspace{-2.5em}
    \caption{Attention mechanism in text-visual alignment. (a) Raw text-visual correlation differences between the NCD and HC groups. (b) Raw position-wise similarities of RoPE. Text-visual attention weights for (c) the HC group (c) and (d) the NCD group, with texts as the query and images as the key. For visualization purposes, the text timeline is resampled (via downsampling or interpolation) to match the number of images.}
    \label{fig:heatmaps}
    \vspace{-1em}
\end{figure}

The Attention-based analysis illustrates how TITAN refines and leverages cross-modal consistency for NCD detection, as illustrated in Fig.~\ref{fig:heatmaps}.
Fig.~\ref{fig:heatmaps}(a) shows the raw cross-modal alignment differences between NCD and HC groups using initial CLIP embeddings. The colormap highlights regions of text-visual correlation divergence, with warm colors indicating higher alignment and cool colors representing lower alignment for the HC group.
Fig.~\ref{fig:heatmaps}(b) displays position-wise cosine similarities from RoPE embeddings, revealing an inherent structure that encodes temporal proximity.

It should be noted that although TITAN operates on raw textual and visual embeddings rather than their correlation maps, the learned Attention weights as shown in Fig.~\ref{fig:heatmaps}(c) and (d) can effectively focus on discriminative text-visual relationships.
Fig.~\ref{fig:heatmaps}(c) shows that for the HC group, Attention highlights temporally coherent regions consistent with RoPE similarities. Fig.~\ref{fig:heatmaps}(d) shows that for the NCD group, Attention shifts towards misaligned or inconsistent cross-modal parts. This adaptive weighting enhances the model's sensitivity to cognitive impairments and thereby improves the performance of NCD detection.

\subsection{Ablation Study of TITAN}

\begin{table}[b]
\centering
\vspace{-1em}
\renewcommand{\arraystretch}{1.05} %
\caption{Ablation study of TITAN by removing each module (``w/o'' means ``without'') in the NCD detection task. The best results are marked in \textcolor{red}{red}, and the second best are \underline{underlined}.}
\label{tab:ablation}
\resizebox{\linewidth}{!}{
\begin{tabular}{lccccc}
\toprule
{\textbf{Approach}} & \textbf{F1} & \textbf{AUC} & \textbf{Rec.} & \textbf{Prec.} & \textbf{Acc.}\\ 
\midrule
TITAN (Sys. 8) & \textcolor{red}{0.7238} & \underline{0.8120} & \underline{0.6333} & \underline{0.8444} & \textcolor{red}{0.7100} \\
w/o RoPE & 0.7171 & \textcolor{red}{0.8203} & 0.6167 & \textcolor{red}{0.8566} & \underline{0.7080} \\
w/o Project-Up\&Down & \underline{0.7218} & 0.7688 & \textcolor{red}{0.6533} & 0.8066 & 0.6980 \\
w/o Attention & 0.6796 & 0.7962 & 0.5833 & 0.8140 & 0.6700 \\
w/o Image emb & 0.6250 & 0.7855 & 0.5000 & 0.8333 & 0.6400 \\
\bottomrule
\end{tabular}
}
\end{table}

We conducted an ablation study to evaluate the contributions of different components of the proposed TITAN model, including \textit{RoPE}, \textit{Project-Up\&Down}, \textit{Attention}, and \textit{Image embeddings}, as detailed in Table~\ref{tab:ablation}. By removing each component (denoted as `w/o'), we assessed its impact on NCD detection performance. The results show that each component is essential, with the absence of the image embedding (w/o Image emb) causing the most significant performance drop, highlighting its critical role in the model's effectiveness.

\subsection{Comparison with previous literature}

To assess model performance, we conducted a comparative evaluation on the public ADReSS and ADReSSo corpora.
As shown in Table~\ref{tab:comparison}, the proposed TITAN model (Sys.8) achieves superior results, with an F1 score of 0.8889 on the ADReSS corpus and 0.8504 on the ADReSSo corpus.

\begin{table}[htb]
\centering
\renewcommand{\arraystretch}{1.05} %
\setlength{\tabcolsep}{5pt} %
\caption{Cross comparison on the ADReSS and ADReSSo corpora.}
\label{tab:comparison}
\begin{tabular}{lccccc}
\toprule
\textbf{Method} & \textbf{F1} & \textbf{AUC} & \textbf{Rec.} & \textbf{Prec.} & \textbf{Acc.} \\ 
\toprule
\multicolumn{6}{c}{ADReSS} \\
\hdashline[2pt/2pt]
Yuan \textit{et. al.}~\cite{yuan2020disfluencies}      & 0.8619 & -      & 0.8540 & 0.8700 & 0.8540 \\ 
Zhu \textit{et. al.}~\cite{zhu2023evaluating} & -      & -      & -      & -      & 0.8224 \\
Wang \textit{et. al.}~\cite{wang2024cross} & 0.8077   & -      & \textcolor{red}{0.8750}     & 0.7500  & 0.7917 \\
Botelho \textit{et. al.}~\cite{botelho2024macro} & 0.8000      & -      & 0.8330      & 0.7690      & 0.8130 \\
Li \textit{et. al.}~\cite{li2024devising}     & 0.8510 & -      & 0.8330 & 0.8700 & 0.8540 \\
Ours (Sys.8)  & \textcolor{red}{0.8889} & \textcolor{red}{0.9410} & 0.8333 & \textcolor{red}{0.9524} & \textcolor{red}{0.8958} \\ 
\toprule
\multicolumn{6}{c}{ADReSSo} \\
\hdashline[2pt/2pt]
Luz \textit{et. al.}~\cite{luz2021detecting} & - & - & - & - & 77.46 \\
P{\'e}rez-Toro \textit{et. al.}~\cite{perez2021influence}  & 0.80 & - & \textcolor{red}{0.8889} & - & 0.8028 \\
Wang \textit{et. al.}~\cite{wang2024cross} & 0.7143 & - & 0.7143 & 0.7143 & 0.7183 \\
Li \textit{et. al.}~\cite{li2024whisper}  & 0.8450 & - & 0.8571 & 0.8333 & 0.8451 \\
Pu \textit{et. al.}~\cite{pu2025integrating} & 0.806 & - & 0.714 & \textcolor{red}{0.926} & 0.831 \\
Shao \textit{et. al.}~\cite{shao2025alzheimer} & 0.8352 & - & 0.8685 & 0.8039 & 0.8315 \\
Ours (Sys.8) & \textcolor{red}{0.8504} & \textcolor{red}{0.9030} & 0.8286 & 0.8742 & \textcolor{red}{0.8564} \\
\bottomrule
\end{tabular}
\vspace{-0.5em}
\end{table}

\section{Conclusion \& Future Work}
\label{sec:conclusion}

Visual-Stimulated Narrative (VSN) serves as a valuable window into integrated cognitive abilities, offering a promising avenue for the automatic and holistic detection of NCDs.
While macrostructural narrative analysis is vital for measuring global coherence for NCD detection, the temporal patterns of topic evolution and cross-modal consistency inherent in VSNs remain underexplored.
To address this, we proposed two dynamic macrostructural modeling approaches -- a DTM-based method and the TITAN model -- to measure the temporal alignment between the narrative topics and the implicit topics present in the visual stimuli. The DTM-based temporal analysis captures the topic evolution of VSNs, while TITAN further models the dynamic topics referenced in sequential images in a fine-grained manner.
These approaches were evaluated on the Cantonese CU-MARVEL-RABBIT corpus, with TITAN achieving superior performance in both classification and regression tasks. Additional cross-comparison on the English ADReSS and ADReSSo corpora, together with feature contribution analysis, further validated the effectiveness of proposed dynamic macrostructural modeling approaches.

We anticipate that the temporal and macrostructural approaches, such as the DTM-based model and TITAN, can significantly contribute to early screening and detection of NCDs, providing a more comprehensive framework for narrative analysis.
Although further technical optimization is possible, the conceptual foundations of these methods offer new insights into quantifying language use associated with cognitive decline, and enhance our understanding of NCDs in VSNs.
Unlike black-box detection models, the proposed methods provide enhanced interpretability through topic evolution and cross-modal alignment analyses, potentially offering clinicians more transparent insights into model decision pathways.
Future work will explore extending these frameworks to incorporate additional modalities, adapting them to multilingual contexts, and applying them to the detection of other speech- and language-related disorders.

\bibliographystyle{IEEEtran}
\bibliography{references}

\appendix
\section{Appendix}
\label{sec:appendix}

\begin{figure}[ht]
    \centering
    \includegraphics[width=\linewidth]{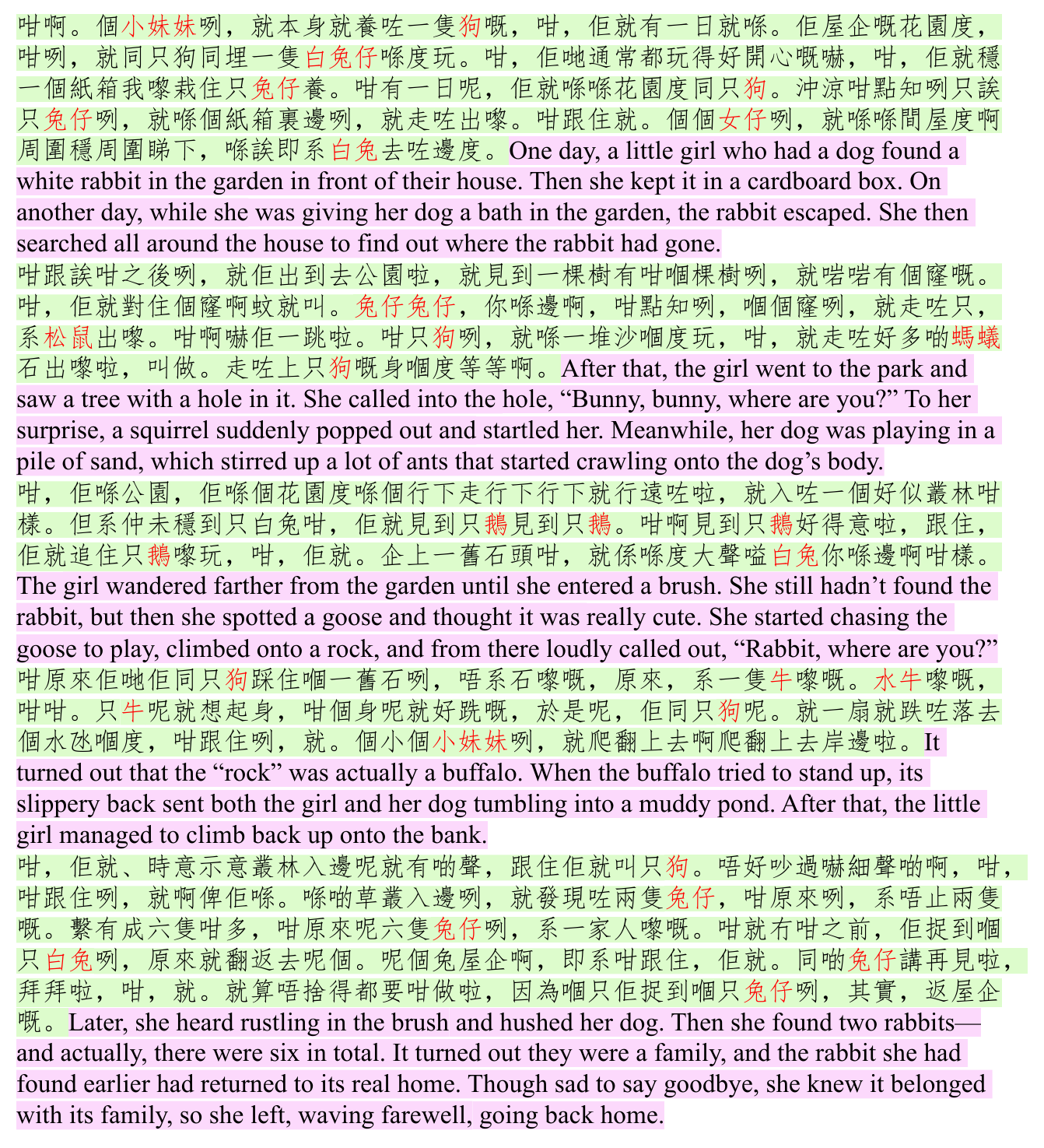}
    \vspace{-2em}
    \caption{Illustrative Cantonese narrative from a healthy control participant in the CU-MARVEL-RABBIT corpus. Rough English translations are provided on a pink background to aid reader comprehension. The characters in the story are highlighted in red color. Normalized topic-related feature values are: consistency = 0.50, cycle = 1.26, ptp range = 1.12, variability = 1.37, change rate = -0.47, temporal correlation = 0.97.}
    \label{fig:hc_example}
\end{figure}

\begin{figure}[t!]
    \centering
    \includegraphics[width=\linewidth]{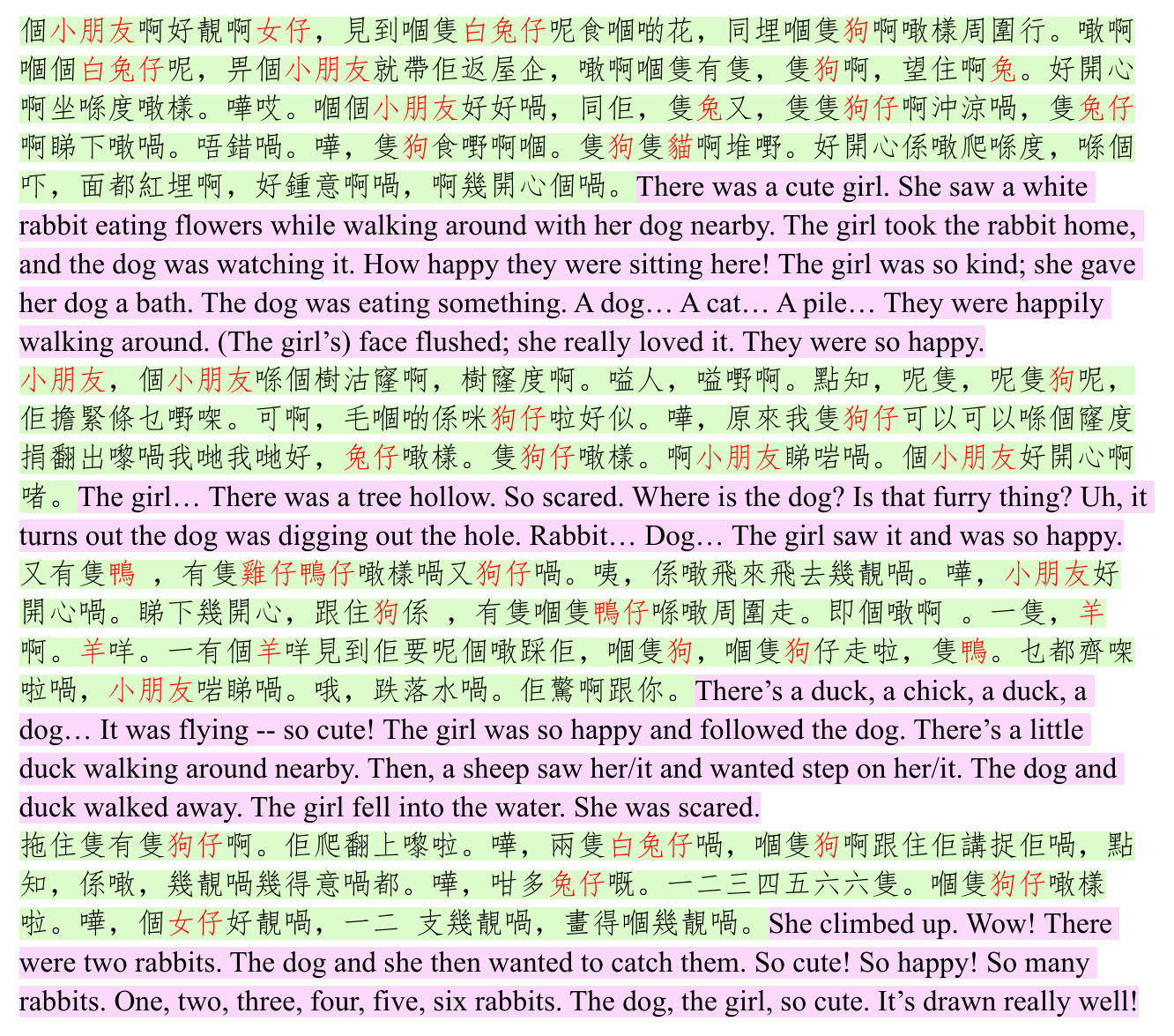}
    \vspace{-2em}
    \caption{Illustrative Cantonese narrative from a participant with NCDs in the CU-MARVEL-RABBIT corpus. Rough English translations are provided on a pink background to aid reader comprehension. The characters in the story are highlighted in red color. Normalized topic-related feature values are: consistency = -0.78, cycle = -0.80, ptp range = -2.94, variability = -2.46, change rate = 0.56, temporal correlation = -1.61.}
    \label{fig:major_ncd_example}
\end{figure}

Full Cantonese narrative transcripts for two exemplar cases mentioned in Section~\ref{sec:data} and Section~\ref{sec:discuss} are provided below. These cases represent a healthy control (Fig.~\ref{fig:hc_example}) and a participant with NCDs (Fig.~\ref{fig:major_ncd_example}), respectively.
Each transcript is accompanied by normalized values of topic-related features (e.g., ``\textit{topic consistency}'', ``\textit{topic cycle}'', ``\textit{topic ptp range}'', ``\textit{topic variability}'', ``\textit{topic change rate}'', ``\textit{topic temporal correlation}'') derived from the proposed DTM method.
To aid reader comprehension, the transcripts are accompanied by rough English translations.

% \begin{textblock*}{\dimexpr0.5\textwidth}[0,0](0.61\textwidth,9.5em)
% \end{textblock*}

% \begin{textblock*}{\dimexpr0.5\textwidth}[0,0](0.61\textwidth,45em)
% \end{textblock*}

\newpage
\vspace{1em}
% Jinchao Li,
\begin{IEEEbiography}[{\vspace{-3.5em}\includegraphics[width=1in,height=1.25in,clip,keepaspectratio]{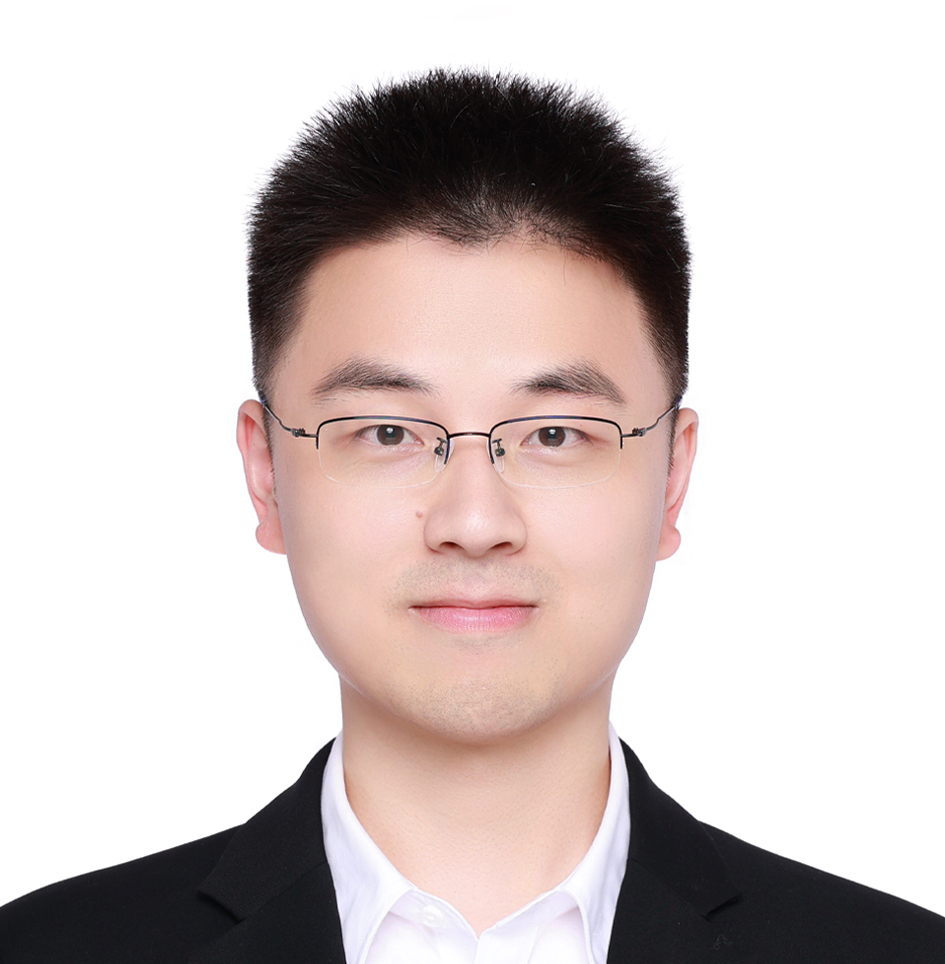}}]{Jinchao Li}
is a Ph.D. candidate at Human-Computer Communications Laboratory (HCCL) in The Chinese University of Hong Kong (CUHK), advised by Prof. Helen Meng. He obtained a B.S. from Nanjing University in 2019. His research focuses on multimodal AI (speech, language, vision) for social good, particularly advancing human-AI interaction for cognitive and mental health applications.
\end{IEEEbiography}
\vspace{-1em}
% \vspace{-5em}

% Yuejiao Wang,
\begin{IEEEbiography}[{\includegraphics[width=1in,height=1.25in,clip,keepaspectratio]{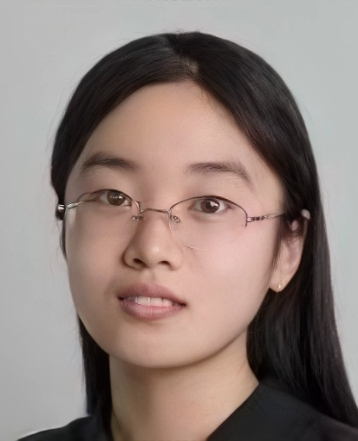}}]{Yuejiao Wang}
is a Ph.D. candidate at HCCL in CUHK, supervised by Professor Helen Meng since August 2021. Her research interests include speech generation, speech recognition, and multimodal healthcare techniques based on fMRI and speech signals. Her work has been published in leading speech-related conferences such as ICASSP and INTERSPEECH. Prior to her Ph.D., she received her M.Sc. degree from the Institute of Automation, Chinese Academy of Sciences, and her B.Eng. degree from Harbin Institute of Technology.
\end{IEEEbiography}
% \vspace{-5em}
% \vspace{2em}

% Junan Li,
\begin{IEEEbiography}[{\includegraphics[width=1in,height=1.25in,clip,keepaspectratio]{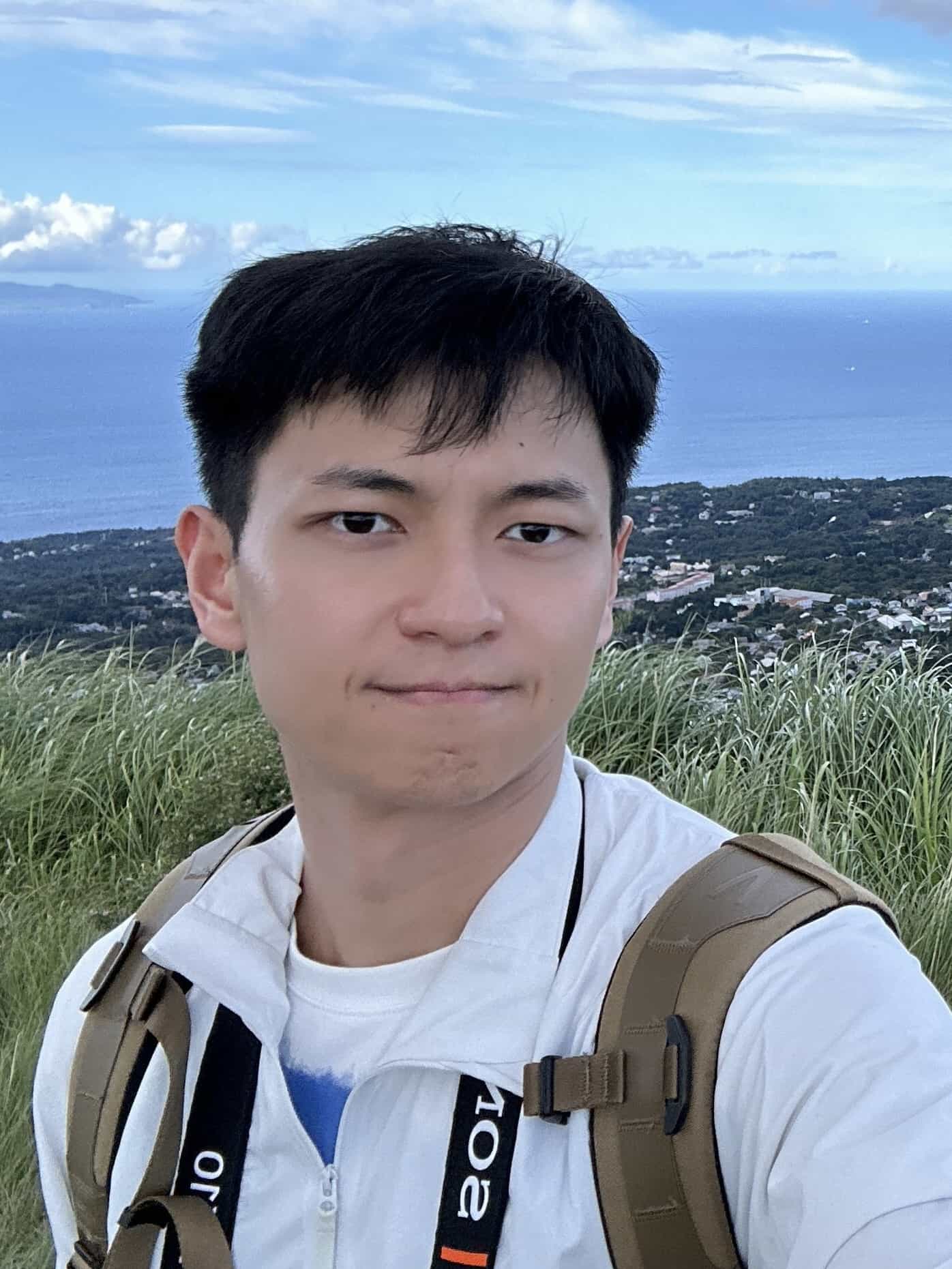}}]{Junan Li}
is a Ph.D. candidate at HCCL in CUHK, under the supervision of Prof. Helen Meng. His research focuses on artificial intelligence for healthcare, particularly on developing explainable spoken language biomarkers for the early detection of neurocognitive disorders. His recent work explores multimodal language understanding by integrating speech, vision, and linguistic representations for clinical applications. He has published in international conferences, and his research interests include natural language processing, multimodal learning, multi-agent systems, and agent planning in AI for medical diagnosis.
\end{IEEEbiography}
% \vspace{-5em}
\vfill

% Jiawen Kang,
\begin{IEEEbiography}[{\includegraphics[width=1in,height=1.25in,clip,keepaspectratio]{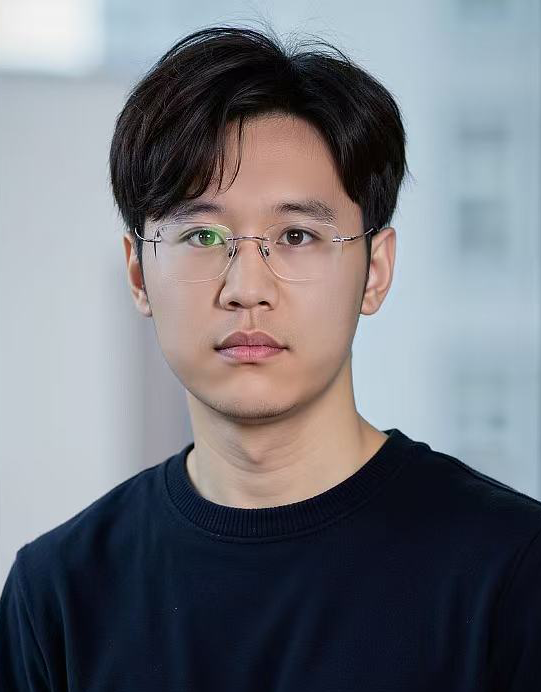}}]{Jiawen Kang}
(Student Member, IEEE) received the B.S. degree from Jilin University in 2019. He worked as Research Assistants at Tsinghua University and CUHK from 2019 to 2021. Since 2021, he has been pursuing a Ph.D. degree with CUHK. His research interests include spoken language processing in conversational diagnostics and conversational speech processing. He received the ISCA Award for the Best Paper Published in Speech Communication in 2024 and the IEEE Ganesh N.Ramaswamy Memorial Student Gran in 2025.
\end{IEEEbiography}
% \vspace{-5em}
\vspace{2em}

% Bo Zheng,
\begin{IEEEbiography}[{\vspace{-4em}\includegraphics[width=1in,height=1in,clip,keepaspectratio]{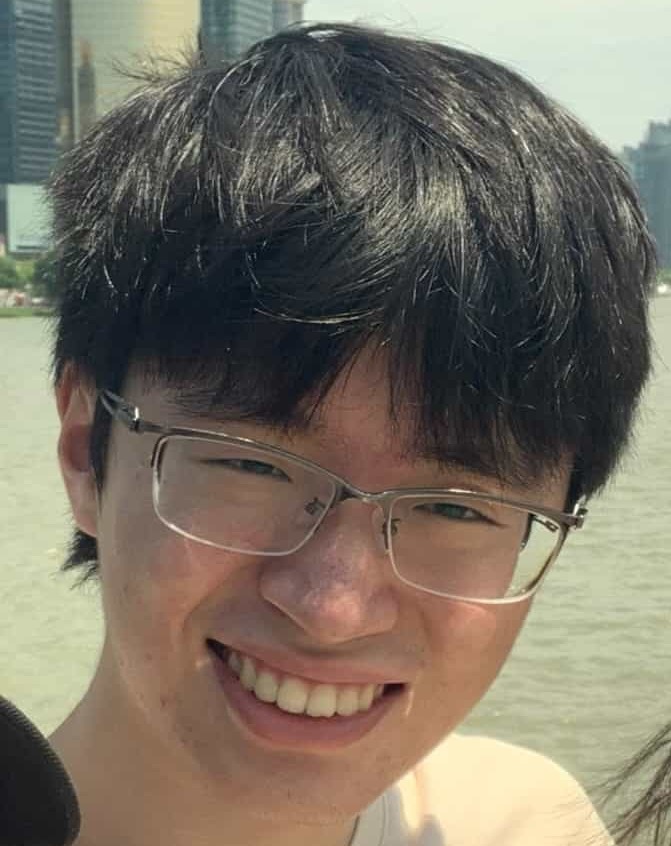}}]{Bo Zheng}
is an M.Phil. student at HCCL in the Dept. of System Engineering and Engineering Management of CUHK, advised by Prof. Helen Meng. His research focuses on the adversarial vulnerability of speech self-supervised learning and the analysis of dysfluency features for neurocognitive disorder detection.
\end{IEEEbiography}
% \vspace{-8em}

% Ka Ho Wong,
\begin{IEEEbiography}[{\includegraphics[width=1in,height=1.25in,clip,keepaspectratio]{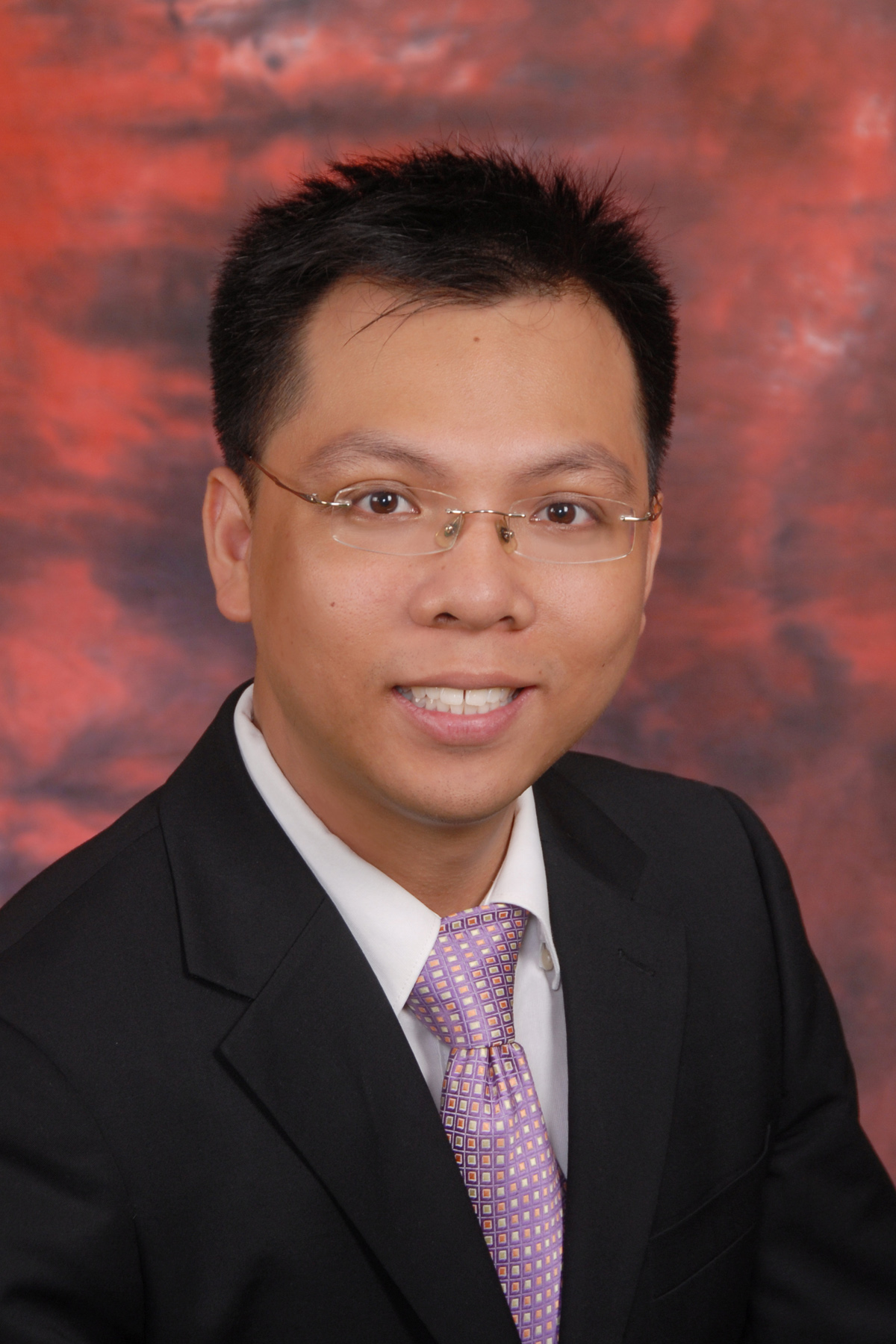}}]{Ka Ho Wong}'s research focuses on dysarthric speech analysis and speech and language processing in neurocognitive disorder. His work aims to understand the characteristics of dysarthric speech and to develop automatic analysis methodologies. His broader research interests include speech technologies, machine learning, visual speech synthesis, e-learning systems and mobile application development. Recently, Dr. Wong has also been actively involved in AI education, speech reconstruction, digital health, and enterprise development.
\end{IEEEbiography}
% \vspace{-5em}
\vspace{2em}

% Brian Mak,
\begin{IEEEbiography}[{\includegraphics[width=1in,height=1.25in,clip,keepaspectratio]{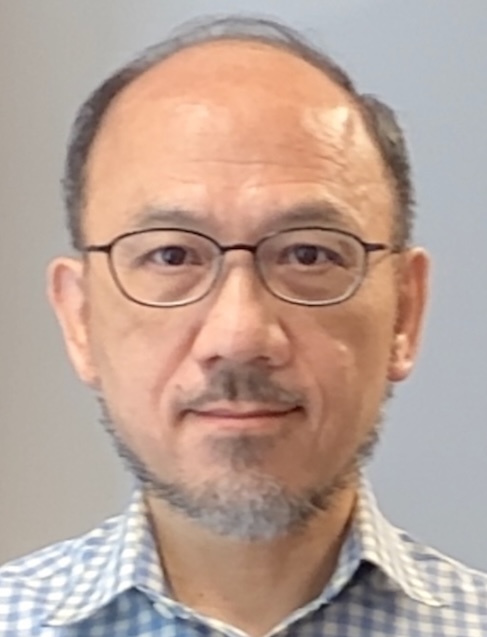}}]{Brian Kan-Wing Mak}
received the B.~Sc. degree in Electrical Engineering from the University of Hong Kong, the M.~S. degree in Computer Science from the University of California, Santa Barbara, USA, and the Ph.D. degree in Computer Science from the Oregon Graduate Institute of Science and Technology, Portland, Oregon, USA.
He worked as a research programmer at the Speech Technology Laboratory of Panasonic Technologies Inc. in Santa Barbara, and as a research consultant at the AT\&T Labs -- Research, Florham Park, New Jersey, USA. 
He was a visiting researcher of Bell Laboratories and Advanced Telecommunication Research Institute -- International as well.
Since April 1998, he has been with the Department of Computer Science \& Engineering in the Hong Kong University of Science and Technology, and is now an Associate Professor.
He has served on the editorial board of the IEEE T-ASLP, Signal Processing Letters, \& Speech Communication; and the Speech \& Language Technical Committee of the IEEE Signal Processing Society.
His interests include acoustic modeling, speech recognition, speech synthesis, voice conversion, sign language recognition and translation, computer-assisted language learning, and machine learning.
He received the Best Paper Award for Speech Processing from the IEEE Signal Processing Society in 2004.
\end{IEEEbiography}
% \vspace{-5em}
\vfill

% Helene H. Fung, https://research.cuhk.edu.hk/en/persons/fung-hoi-lam-helene
\begin{IEEEbiography}[{\includegraphics[width=1in,height=1.25in,clip,keepaspectratio]{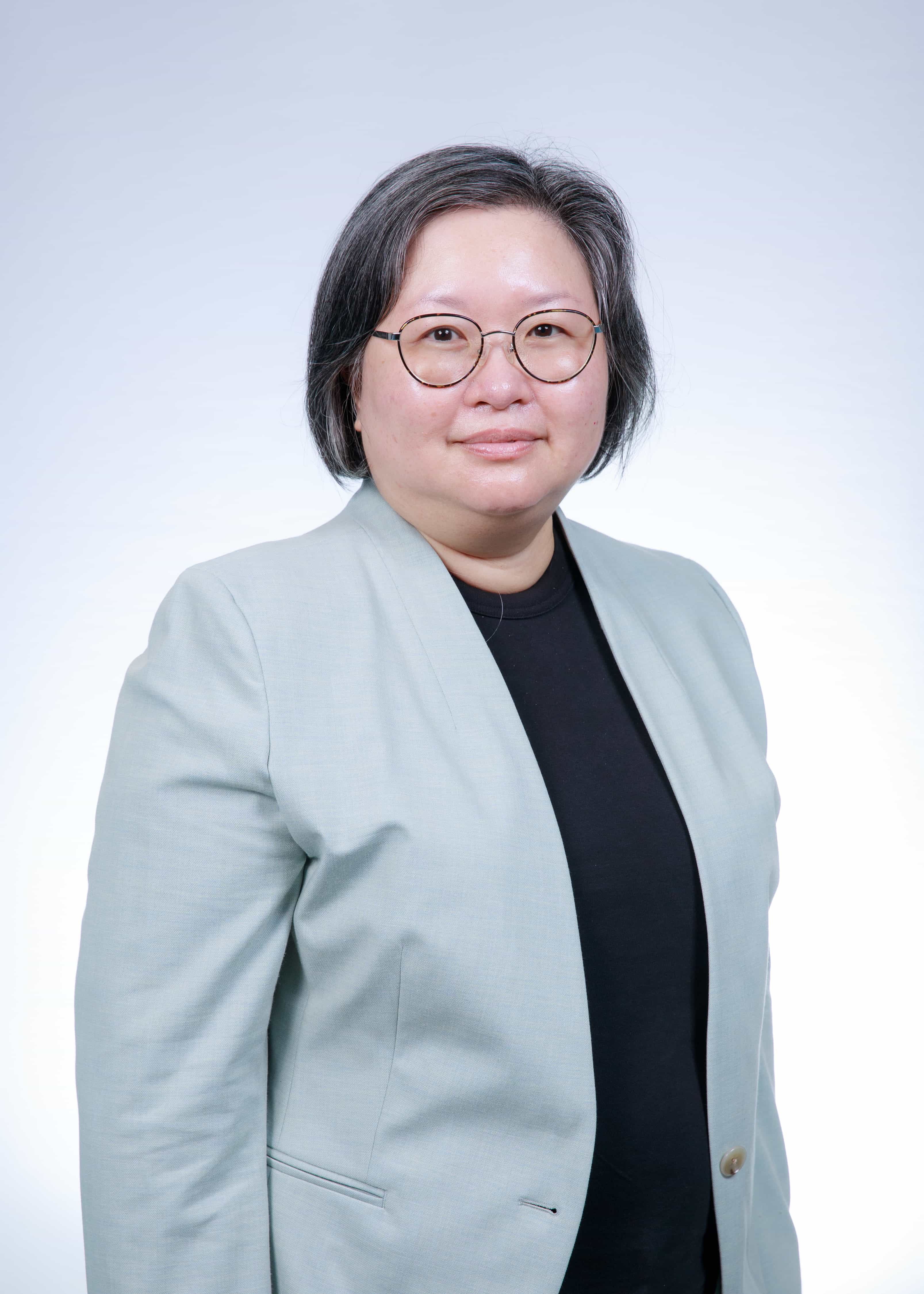}}]{Helene H. Fung}
is a Professor at the Dept. of Psychology, the Executive Director of the Centre for Positive Social Science and a Deputy Director of the Institute of Ageing, at CUHK. She was an assistant professor at the University of Alberta, Canada. She obtained her B.S. from University of Washington, Seattle, and MA and PhD from Stanford University. Professor Fung examines socioemotional ageing across cultures.
Among her awards include the 2024 Healthy Longevity Catalyst Award from the US National Academy of Medicine, 2017-18 Anne E. Ofstedal Fellowship on Higher Education Leadership, offered by United Board, the 2010 Margret and Paul Baltes award in Behavioral and Social Gerontology from the Gerontological Society of America, the 2008 Retirement Research Foundation Mentor Award from Division 20 and the 2016 Henry David International Mentoring Award from Division 52, American Psychological Association.
She is included on a list of the top 2\% of working scientists world-wide and is ranked among Top Scientists in Psychology in China by Research.com. She is a senior associate editor for the Australian Journal of Psychology, and an associate editor for Cognition and Emotion. She has been an associate editor for Acta Psychologica Sinica. She was elected a fellow of the Association for the Psychological Science, a fellow (Divisions 20 \& 52) and member-at-large (Division 20) of the American Psychological Association (Division 20 \& 52), and a fellow and Member-at-large of the Behavioral and Social Science Division of the Gerontological Society of America.
Her research focuses on how goals change across adulthood, and their impacts on social relationships, emotional regulation, and cognition.
\end{IEEEbiography}
\vspace{-5em}

% Jean Woo, https://research.cuhk.edu.hk/en/persons/jean-woo
\begin{IEEEbiography}[{\includegraphics[width=1in,height=1.25in,clip,keepaspectratio]{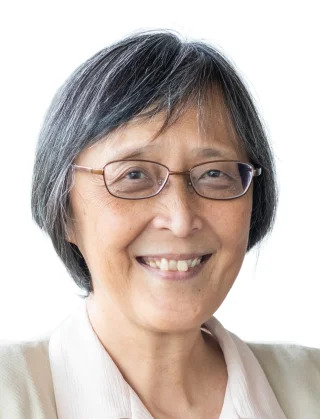}}]{Jean Woo}
graduated from the University of Cambridge in 1974. After medical posts in the Charing Cross, Hammersmith, and Brompton Hospitals in the UK, she worked in part time posts in general practice as well as research at the University of Hong Kong.
Prof. Woo joined the Dept. of Medicine at CUHK in 1985 as Lecturer responsible for the development of the teaching and service in Geriatric Medicine, becoming Head of the Dept. in 1993 until 1999, Chief of Service of the Medicine and Geriatric Unit at Shatin Hospital from 1993 to 2012, and Chair Professor of Medicine in 1994.
She served as Head of the Dept. of Community and Family Medicine (2000–2006), Director of the School of Public Health (2001–2005), and Chairman of the Dept. of Medicine \& Therapeutics (2013–2016).
She established the Centre for Nutritional Studies in 1997 using a self financing model to carry out service, education and research; and the Centre for Gerontology and Geriatrics in 1998, offering self-financed courses in Gerontology and Geriatrics, as well as End of Life Care.
Currently she is the Co-Director of CUHK Institute of Health Equity, Director of the Jockey Club Institute of Aging at CUHK, and Honorary Consultant of the Prince of Wales and Shatin Hospitals, Hospital Authority.
Her research interests include epidemiology, public health, health and social care systems aspects of aging, integrated primary care models for older adults, health equity for aging populations, etc.
\end{IEEEbiography}
\vspace{-5em}

% Man-Wai Mak,
\begin{IEEEbiography}[{\includegraphics[width=1in,height=1.25in,clip,keepaspectratio]{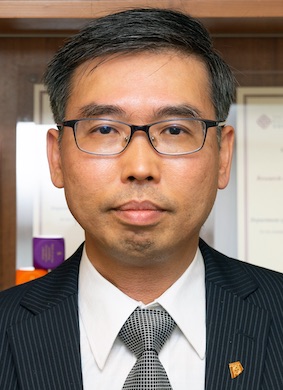}}]{Man-wai MAK}
(Senior Member, IEEE) received a BEng and PhD in Electronic Engineering from Newcastle Upon Tyne Polytechnic and Northumbria University in 1989 and 1993, respectively.
He joined at The Hong Kong Polytechnic University in 1993, served as Interim Head of the Dept. of Electronic and Information Engineering from 2021 to 2023, and Professor of the Dept. of Electrical and Electronic Engineering from 2024.
He published over 250 articles and books and co-authored the postgraduate textbooks ``Biometric Authentication: A Machine Learning Approach, Prentice-Hall, 2005,'' and ``Machine Learning for Speaker Recognition, Cambridge University Press, 2020.''
He has received three Faculty of Engineering Research Grant Achievement Awards and a Faculty Award for Outstanding Performance.
He served as Chairman of the IEEE Hong Kong Section Computer Chapter (2003–2005), member of the IEEE Machine Learning for Signal Processing Technical Committee (2005–2007), Area Chair of Interspeech 2014 and ICTAI 2016, and ISCSLP Steering Committee member (2014–2018) and Program Co-Chair (2018, 2021).
His research interests include speaker recognition, machine learning, spoken language processing, biomedical engineering, and bioinformatics.
\end{IEEEbiography}
% \vspace{-4em}

% Timothy Kwok, https://research.cuhk.edu.hk/en/persons/kwok-chi-yui-timothy
\begin{IEEEbiography}[{\includegraphics[width=1in,height=1.25in,clip,keepaspectratio]{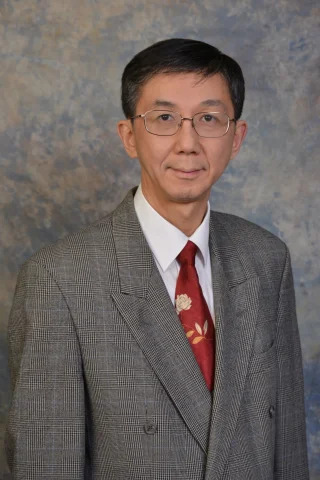}}]{Timothy Kwok}
received specialist training in Geriatric and Internal Medicine in St George’s Hospital, London. He joined the Dept. of Medicine \& Therapeutics at CUHK in 1994, became professor in 2006, and was endowed SC Ho Professorship in 2022. He is director of Jockey Club Centre for Positive Ageing and Jockey Club centre for osteoporosis care and control.
His research focuses on the prevention strategies and non-pharmacological interventions for dementia. Since 2015, he has set up a prospective cohort of over 1200 older people with Alzheimer disease (AD) and mild cognitive impairment to examine the determinants of functional decline in AD.
He also studies fracture prevention and healthy ageing. Since 2001, he has led a prospective cohort study of 4,000 older men and women, examining risks for osteoporosis, fractures, cognitive decline, and more.
% The large data set has shown insights in the risk factors of osteoporosis, fractures, frailty, sarcopenia, functional decline, cognitive decline, atherosclerosis and mortality. In more recent follow-up visits, wrist actigraphy, voice, metabolomics, gut microbiota, brain and spine MRI have been added.
His research interests include Dementia, nutrition in older people, nutritional supplements for brain health, Osteoporosis, and fall prevention.
\end{IEEEbiography}
% \vspace{-4em}

% Vincent Mok, https://research.cuhk.edu.hk/en/persons/mok-chung-tong-vincent
\begin{IEEEbiography}[{\includegraphics[width=1in,height=1.25in,clip,keepaspectratio]{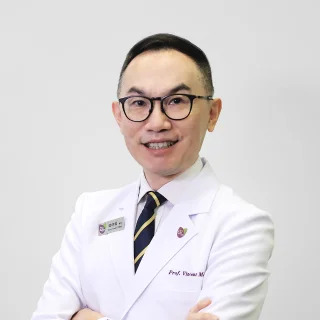}}]{Vincent Mok}'s research focuses on understanding the mechanisms of dementia and investigating strategies for dementia prevention.
He was the first to report the association between cerebral small vessel disease (CSVD) and cognitive impairment in Asia, and its community prevalence among Asians (Chinese).
% These findings have increased awareness that strategies targeting CSVD, such as blood pressure lowering, can help prevent dementia in this region with a particularly high dementia burden.
Prof. Mok led one of the largest single-centre studies on cognitive decline in the context of stroke.
He found that the concurrent presence of amyloid plaques, detected by in-vivo amyloid PET imaging, and CSVD are important factors associated with early and delayed cognitive decline after stroke, respectively.
% These findings provide a roadmap for preventing vascular cognitive impairment.
Prof. Mok, along with Dr. Adrian Wong, validated the HK-MoCA, and established age- and education-adjusted norms to support its widespread use in Hong Kong.
% HK-MoCA is currently the most commonly used brief cognitive assessment for evaluating cognitive function and detecting mild cognitive impairment in Hong Kong.
His research focuses on mechanisms and therapeutics of common age related cognitive disorders and Parkinson’s disease, as well as AI-aided imaging technologies for the diagnosis and monitoring of brain diseases.
% research interests
% -Mechanisms and therapeutics of common age related cognitive disorders, e.g., Alzheimer’s disease, Vascular Cognitive Impairment, Cerebral Small Vessel Disease, Mixed Dementia
% -Artificial intelligence aided technology based on MRI and retinal imaging for diagnosis, prognostication, and monitoring of brain diseases
% -Parkinson's disease - premotor stage, genetics, neuroimaging, and Deep Brain Stimulation
\end{IEEEbiography}
% \vspace{-4em}

% Xianmin Gong,
\begin{IEEEbiography}[{\vspace{-2.5em}\includegraphics[width=1in,height=1.25in,clip,keepaspectratio]{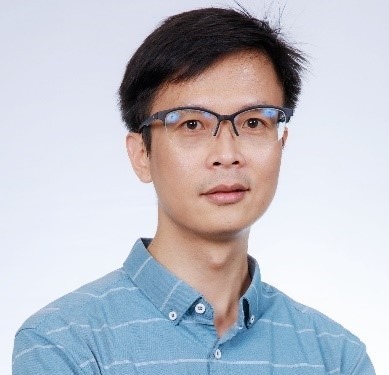}}]{Xianmin Gong}
obtained his Ph.D. in psychology from CUHK and received postdoctoral training at the University of Zurich, Switzerland. He worked as a research assistant professor at the BDDA Research Center at CUHK before joining the Dept. of Psychology as a lecturer. His research focuses on motivational development, well-being across adulthood, and cognitive aging.
\end{IEEEbiography}
% \vspace{-4em}

% Xixin Wu,
\begin{IEEEbiography}[{\vspace{1em}\includegraphics[width=1in,height=1.25in,clip,keepaspectratio]{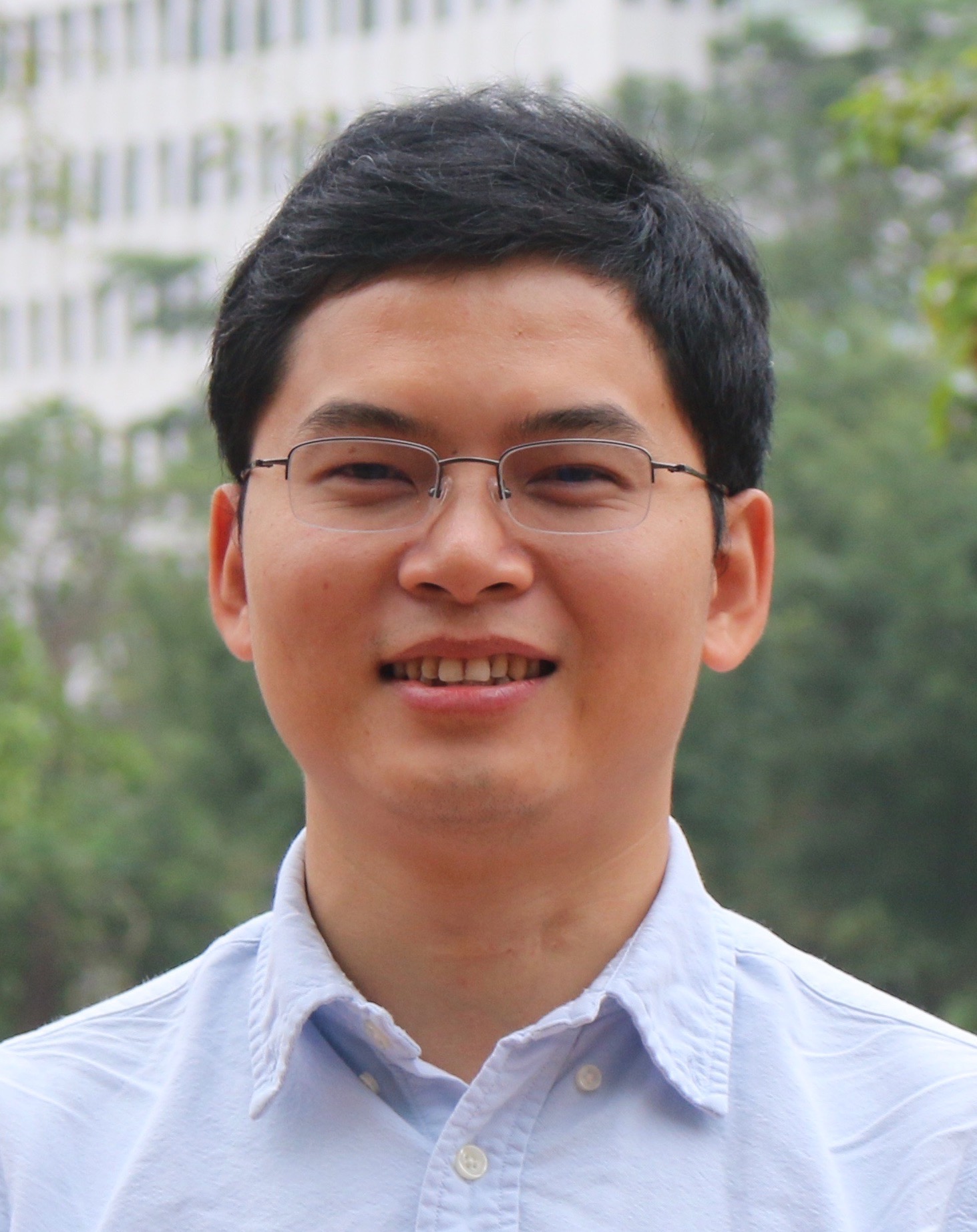}}]{Xixin Wu}
(Member, IEEE \& ISCA) received his B.S. from Beihang University, Beijing, China, M.S. from Tsinghua University, Beijing, China, and Ph.D. from CUHK, Hong Kong. He was a Research Associate at the Machine Intelligence Laboratory, Cambridge University Engineering Dept.. Since 2021, he has been a Research Assistant Professor at the Stanley Ho Big Data Decision Analytics Research Centre, CUHK. His research interests include speech synthesis and recognition, speaker verification, and neural network uncertainty.
\end{IEEEbiography}
% \vspace{-5em}

% Xunying Liu,
\begin{IEEEbiography}[{\includegraphics[width=1in,height=1.25in,clip,keepaspectratio]{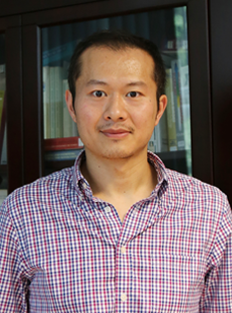}}]{Xunying Liu}
(Member, IEEE \& ISCA) received his B.S. degree from Shanghai Jiao Tong University, Shanghai, China, and M.Phil. degree in computer speech and language processing and Ph.D. degree in speech recognition from the University of Cambridge, Cambridge, U.K. He had been a Senior Research Associate with the Machine Intelligence Laboratory, Dept. of Engineering, University of Cambridge, and since 2016 has been an Associate Professor with the Dept. of Systems Engineering and Engineering Management, CUHK, Hong Kong. His research interests include machine learning, large vocabulary continuous speech recognition, statistical language modelling, noise robust speech recognition, audio-visual speech recognition, computer aided language learning, speech synthesis and assistive technology. He and his students received a number of best paper awards and nominations, including best paper awards at ISCA Interspeech 2010 and IEEE ICASSP 2019.
% including a Best Paper Award at ISCA Interspeech2010 for the paper titled “Language Model Cross Adaptation for LVCSR System Combination”, and Best Paper Award at IEEE ICASSP2019 for their paper titled “BLHUC: Bayesian Learning of Hidden Unit Contributions for Deep Neural Network Speaker Adaptation”.
\end{IEEEbiography}
% \vspace{-5em}

% Patrick C. M. Wong, https://research.cuhk.edu.hk/en/persons/wong-patrick-chun-man
\begin{IEEEbiography}[{\includegraphics[width=1in,height=1.25in,clip,keepaspectratio]{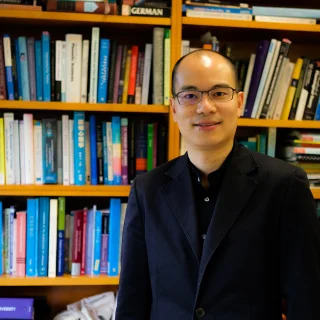}}]{Patrick C. M. Wong}
is Professor of Cognitive Neuroscience and Linguistics and Director of Brain and Mind Institute at The Chinese University of Hong Kong. His research covers a wide range of basic and translational issues concerning the neural basis and disorders of language and music. His research has been funded by the National Science Foundation (USA), the National Institutes of Health (USA), and the Research Grants Council of Hong Kong, including an Areas of Excellence award. He is a Guggenheim Fellow.

\end{IEEEbiography}
% \vspace{-5em}

% Helen Meng,
\begin{IEEEbiography}[{\includegraphics[width=1in,height=1.25in,clip,keepaspectratio]{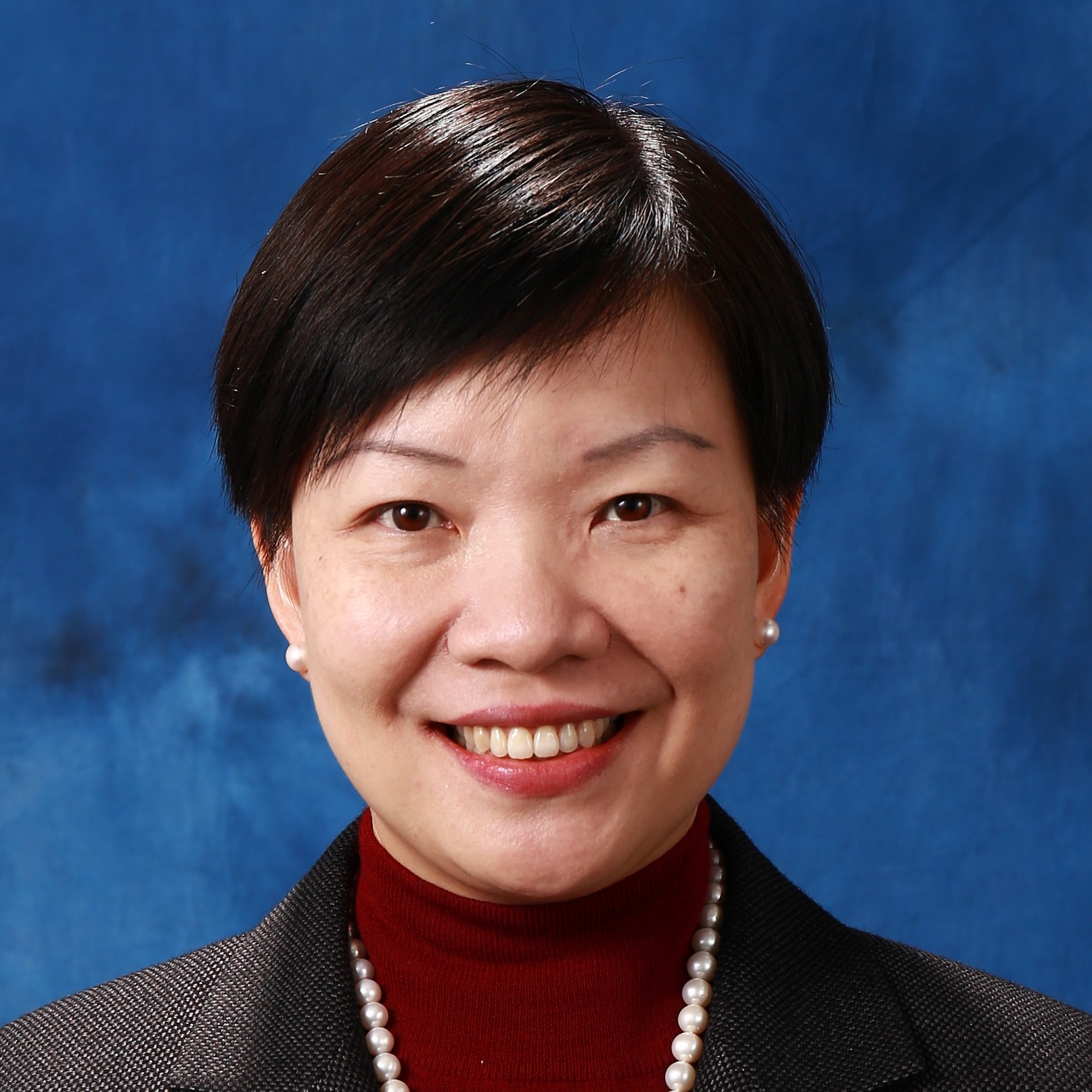}}]{Helen Meng}
(Fellow, IEEE) received the B.S., M.S., and Ph.D. degrees in electrical engineering from the Massachusetts Institute of Technology, Cambridge, MA, USA. In 1998, she joined the Chinese University of Hong Kong, Hong Kong, where she is currently a Patrick Huen Wing Ming Professor with the Dept. of Systems Engineering \& Engineering Management. She was the former Dept. Chairman and the Associate Dean of Research with the faculty of Engineering. Her research interests include human-computer interaction via multimodal and multilingual spoken language systems, spoken dialog systems, computer-aided pronunciation training, speech processing in assistive technologies, health-related applications, and big data decision analytics. She was the Editor-in-Chief of the IEEE T-ASLP between 2009 and 2011. She was the recipient of the IEEE Signal Processing Society Leo L. Beranek Meritorious Service Award in 2019. She was also an Elected Board Member of the International Speech Communication Association (ISCA) and an ISCA International Advisory Board Member. She is a Fellow of ISCA, HKCS, HKIE and IEEE.
\end{IEEEbiography}

%\begin{IEEEbiography}[{\includegraphics[width=1in,height=1.25in,clip,keepaspectratio]{mshell}}]{Michael Shell}\end{IEEEbiography} \vspace{-5em}
% \end{IEEEbiography} \vspace{-5em}

% \begin{IEEEbiography}[{\includegraphics[width=1in,height=1.25in,clip,keepaspectratio]{xxx}}]{Michael Shell}
% \end{IEEEbiography} \vspace{-5em}

%\newpage


\vfill

% \enlargethispage{-5in}

\end{document}